\documentclass[%
 aip,
 amsmath,amssymb,
preprint,%
]{revtex4-1}

\usepackage{graphicx}
\usepackage{dcolumn}
\usepackage{bm}

\usepackage{epstopdf, epsfig}
\usepackage{subfigure}
\usepackage{multirow}
\usepackage{mathrsfs}
\usepackage{ esint }

\usepackage[T1]{fontenc}
\usepackage{mathptmx}

\begin{document}

\preprint{AIP/123-QED}

\title[Sample title]{Two-stage growth mode for lift-off mechanism in oblique shock-wave/jet interaction}

\author{Bin Yu}
\author{Miaosheng He }
\author{Bin Zhang }
\author{Hong Liu}
\email{hongliu@sjtu.edu.cn}
\affiliation{School of Aeronautics and Astronautics, Shanghai Jiao Tong University}

\date{\today}

\begin{abstract}
The lift-off characteristics of supersonic streamwise vortex in oblique shock-wave/jet interaction (OS/JI for short), extracted from a wall-mount ramp injector in scramjet, is studied through Large-eddy simulation method.
  Shocked helium jet is deformed into a pair of streamwise vortex with a co-rotating companion vortex showing the lift-off characteristic immediately after shock.
  Based on the objective coordinate system in frame of oblique shock structure, it is found that the nature of three-dimensional lift-off structure of a shock-induced streamwise vortex is inherently and precisely controlled by a two-stage growth mode of structure kinetics of a shock bubble interaction (SBI for short).
  The striking similar structures between OS/JI and SBI support the proposition that the lift-off of streamwise vortex is the result of a underlying two-dimensional vortical motion.
  By considering the first stage impulsive linear growth rate, an improved vortex propagation model suitable for SBI is proposed and validated.
  The lift-off phenomena of both numerical OS/JI case in this paper and wall-mounted ramp injector cases in literature are well explained under the two-stage structure kinetics model of SBI.
  This model further predicts that for higher free stream Mach number ($M>5$), increasing ramp compression shows little effect on elevating streamwise vortex while evident lift-off may occur for lower Mach number ($M<3.5$), which offers the new way for preliminary design of streamwise vortex-based ramp injector in scramjet.
\end{abstract}

\maketitle

\section{Introduction}
\label{sec: intro}
Fuel-air mixing enhancement of supersonic internal flows such as in scramjet combustor is the notorious challenging problem because the mixing strategy developed in subsonic flows, for example parallel shear mixing \cite{dimotakis1991turbulent}, loses its strength and can not be easily extended to supersonic flows \cite{curran1996fluid} due to the intrinsic compressible effect \cite{gutmark1995mixing}.
Swithebank \cite{Swithebank1969Vortex} firstly proposed the streamwise vortex generation from the swirl jet to enhance mixing in supersonic flows.
Although the device of swirler introduced in supersonic flows will lead to the large total pressure loss~\cite{seiner2001historical}, the idea of the supersonic streamwise vortex enhancing mixing is gradually the aim in scramjet combustor~\cite{maddalena2014vortex}. Recently, a systematic review on several scramjet model combustors that are equipped with different kinds of mixing strategies addresses that streamwise vortex is still widely accepted as one of the most effective way to enhance mixing \cite{urzay2018supersonic}.

Among all descriptors concerned in streamwise vortex enhanced mixing, lift-off characteristic, or say penetration trajectory of fuel injection, is one of the important parameters that determines the injector performance \cite{ben2006time}.
For one reason, if the fuel can be controlled to penetrate into the center of main flow to the greatest extent, the contact area between fuel and air is largely expanded, which further increases mixing \cite{lee2006characteristics}.
For another reason, if combustion is considered, well-posed penetration of fuel can alleviate the heat destruction on the combustor wall \cite{gamba2015ignition}. Moreover, proper heat release distribution, which seriously relies on lift-off of fuel jet, is the key to a successful mode-transition in dual-mode ramjet/scramjet combustors \cite{aguilera2017scramjet}. Therefore, lift-off of fuel injection from the supersonic streamwise vortex is the core evaluation in the design of combustors.

A lateral injection in crossflow can form a pair of lifting counter-rotating vortex, which is a canonical problem in mixing enhancement.
As for incompressible flows, the penetration trajectory of lateral injection is mainly controlled by normal momentum flux equivalent to a normal force \cite{broadwell1984structure}. When supersonic inflow is considered, compressibility becomes important. Jet penetration and trajectory can be dependent not only on momentum flux ratio \cite{papamoschou1993visual}, but Mach number, molecular weights or geometric shape of orifice \cite{Mahesh2013The}.
Insight mechanism of jet in crossflow has been reviewed extensively in Refs.~\onlinecite{karagozian2010transverse,Mahesh2013The,karagozian2014jet}.
However, a normal injection of gaseous fuel may lose its effect at relatively high inflow Mach number as Ma$>3$ due to the unaffordable injection pressure, which indicates that a liquid-fuel injection is preferred \cite{lee2015challenges}.
Thus, producing streamwise vortex other than normal injection to enhance gaseous mixing at hypervelocity is desired \cite{hiejima2020shockwave}.
Utilizing the potential shock structures embedded in supersonic flows, the so-called shock-enhanced mixing mechanism is firstly introduced by Marble \cite{Marble1989Progress,marble1994gasdynamic}. Artificial or natural oblique shock wave is set to interact with fuel jet, which creates strong axial vortices that stretches the fuel-air interface \cite{curran1996fluid}.
Oblique shock and circular jet interaction is widely-applied in the research of  scramjet or shramjet (shock induced combustion ramjet~\cite{sislian2006numerical}).
Classic wall-mounted ramp injector is raised by Waitz \emph{et al} \cite{waitz1993investigation} and following studied by numerical method in Ref.~\onlinecite{lee1997computational}.
This kind of injector compressed the supersonic flow at downedge, which forms a oblique shock interacting with jet injected from ramp.
The maximum distortion and life-off of jet from oblique shock interaction is realized \cite{parent2004hypersonic}. Faster mixing characteristic from oblique shock/jet interactions shows its potential in reducing combustor length in hypersonic internal flows~\cite{landsberg2018enhanced}.
Although wall-mounted injector shows the desirable results of mixing enhancement from shock interaction, the underlying mechanism of lift-off characteristic in this type of shock-induced supersonic streamwise vortex, which is clearly different from jet in crossflow, remains to be revealed.

The fundamental mechanism of streamwise vortex benefit from shock is the baroclinic vorticity production that is firstly found in Richtmyer-Meshkov instability (RMI for short) phenomenon \cite{richtmyer1960taylor,meshkov1969instability}.  Under the slender body approximation \cite{yang1993applications,yang1994model}, a three-dimensional steady oblique shock jet interaction (OS/JI for short) can be analogous to two dimensional shock bubble interaction (SBI for short),  a canonical case in the vigorous research of RMI \cite{brouillette2002richtmyer,ranjan2011shock}.
By this way, the lift-off structure can be further studied in view of shock bubble interaction.
However, the slender body approximation proposed in Ref.~\onlinecite{yang1994model} is only qualitatively proven, which shows large deviation from the wall-mounted injector.
This deviation is further confirmed in present paper.
Thus, there exists a transformation gap between the scientific research in RMI and the real wall-mounted injector, which is vital to understand lift-off phenomena and its mechanism for shock-induced supersonic streamwise vortex.

To investigate lift-off phenomena in oblique shock/jet interaction from a wall-mounted ramp injector, the structures of streamwise vortex in the typical condition of internal flow in scramjet with Ma$_{3D}$=3.5 are numerically studied.
Based on the objective coordinate system on oblique shock, we will quantitatively prove that the lift-off structures of oblique shock jet interactions are inherent and precisely controlled by two-stage structure kinetics of an analogous counterpart of shock bubble interaction.
A modified two-stage structure kinetics model well explains the lift-off phenomena in both oblique shock jet interaction concerned in this paper and experimental/numerical ramp injector in literature.
Our results provide the insight into the nature of lift-off phenomena in oblique shock wave/jet interaction in the novel view of dynamics in shock bubble interaction and builds a vigorous connection between the basic scientific research in RMI and the real wall-mounted injector.

The remainder of this paper is organized as follows. In Sec.\ref{sec: numer}, numerical method and initial conditions for case studied in this paper are described. Sec.\ref{sec: 3DJSI} shows the results of the lift-off phenomena in oblique shock wave interaction. The analogy of lift-off structure to structure kinetics of shock bubble interaction is proven in Sec.\ref{sec: 2D3D}.
Sec.\ref{sec: 2D SBI} investigates the vortex kinetics in shock bubble interaction, which leads to a physical model of lift-off for oblique shock jet interaction shown in Sec.\ref{sec: liftoff}.
Conclusions and suggestions for future work are presented in Sec.\ref{sec: conclu}.

\section{Numerical method and setup}
\label{sec: numer}
\subsection{Numerical method}
In this paper, three-dimensional Large eddy simulation is performed by our in-house code platform \emph{ParNS3D} \cite{wang2018scaling,liang2019hidden,liu2020optimal} to study the oblique shock jet interaction.
The governing equations of mass, momentum, energy and transportation of species are:
\begin{equation}\label{eq: masscon}
  \frac{\partial\overline{\rho}}{\partial t}+\frac{\partial (\overline{\rho}\widetilde{u}_i)}{\partial x_i}=0
\end{equation}
\begin{equation}\label{eq: momcon}
  \frac{\partial (\overline{\rho}\widetilde{u}_i)}{\partial t}+\frac{\partial(\overline{\rho}\widetilde{u}_i\widetilde{u}_j)}{\partial x_j}=\frac{\partial\left(\widetilde{\sigma}_{ij}-\widetilde{\tau}_{ij}\right)}{\partial x_j}-\frac{\partial\widetilde{p}}{\partial x_i}
\end{equation}
\begin{equation}\label{eq: enercon}
  \frac{\partial(\overline{\rho}\widetilde{E})}{\partial t}+\frac{\partial(\overline{\rho}\widetilde{u}_i\widetilde{H})}{\partial x_i}=\frac{\partial\left(\widetilde{u}_j\left(\widetilde{\sigma}_{ij}-\widetilde{\tau}_{ij}\right)\right)}{\partial x_i}-\frac{\partial (\widetilde{q}_i+\widetilde{Q}_i)}{\partial x_i}
\end{equation}
\begin{equation}\label{eq: mfcon}
  \frac{\partial(\overline{\rho}\widetilde{Y}_s)}{\partial t}+\frac{\partial(\overline{\rho}\widetilde{u}_i\widetilde{Y}_s)}{\partial x_i}=\frac{\partial}{\partial x_i}\left(\overline{\rho}\mathcal{D}\frac{\partial\widetilde{Y}_s}{\partial x_i}-\widetilde{b}_{is}\right)
  \qquad s=1,2,\cdots,m-1
\end{equation}
where $\overline{(\cdot)}$ is Reynolds averaging and $\widetilde{(\cdot)}$ is the Favre filtered averaging.
All variables are decomposed to resolved and unresolved (subgrid) parts by spatial filters that $f=\widetilde{f}+f'$ with $\widetilde{f}=\overline{\rho f}/\overline{\rho}$.
Moreover, $\overline{\rho}$, $\widetilde{u}_j$, $\widetilde{p}$, $\widetilde{E}$ and $\widetilde{H}$ is density, velocity, pressure, energy and enthalpy respectively.
The mass fraction of $sth$ component is denoted as $\widetilde{Y}_s$.

Also in above equations, $\widetilde{\sigma}_{ij}=\mu[\partial\tilde{u}_i/\partial x_j+\partial\tilde{u}_j/\partial x_i-2/3\delta_{ij}\partial\tilde{u}_k/\partial x_k]$ is the viscous stress tensor, $\widetilde{\tau}_{ij}\equiv\overline{\rho}\widetilde{u_iu_j}-\overline{\rho}\tilde{u}_i\tilde{u}_j$ is the subgrid scale stress tensor (SGS), $\widetilde{q}_i=-\lambda\partial\widetilde{T}/\partial x_i$ is the heat flux where $\widetilde{T}$ is temperature, $\widetilde{Q}_i\equiv\overline{\rho}\widetilde{u_iH}-\overline{\rho}\tilde{u}_i\widetilde{H}$ is the subgrid heat flux and $\widetilde{b}_{is}\equiv\overline{\rho}\widetilde{u_iY_s}-\overline{\rho}\tilde{u}_i\widetilde{Y}_s$ is subgrid diffusion term \citep{sabelnikov2013combustion}.

As for component transport coefficient, $\mu$, $\lambda$ is the dynamic viscosity and thermal conductivity of the mixed gas, given by the Wike's semi-empirical formula \cite{Gupta1989A}. $\lambda=C_p\mu/\mathrm{Pr}$ in which $C_p$ is constant pressure specific heat and Prandtl number is chosen as $\mathrm{Pr}=0.72$  \cite{houim2011low}.
For high-speed flow numerical simulations, mass diffusion $\mathcal{D}$ can be simplified by ignoring pressure and temperature diffusion, and its assumed to be constant with different components as $\mathcal{D}_i=\mathcal{D}=\mu/(\rho \mathrm{Sc})$ where Schmidt number is constant as $\mathrm{Sc}=0.5$ \cite{Gupta1989A}.

For the compressible flows and under Boussinesq hypothesis \citep{hinze1975turbulence}, the subgrid scale stress tensor can be modelled as $\widetilde{\tau}_{ij}=-2\mu_t(\widetilde{S}_{ij}-1/3\widetilde{S}_{kk}\delta_{ij})$, where $\widetilde{S}_{ij}=1/2(\partial\tilde{u}_i/\partial x_j+\partial\tilde{u}_j/\partial x_i)$ is strain rate tensor and $\mu_t$ is turbulent eddy-viscosity. Subgrid heat flux can be modelled as $\widetilde{Q}_i=-\lambda_t\partial\widetilde{T}/\partial x_i$, where $\lambda_t=\mu_tC_p/Pr_t$ is turbulent heat conductivity and $Pr_t=0.9$ is turbulent Prantl number. Subgrid diffusion flux can be modelled as $\widetilde{b}_{is}=-\overline{\rho}\mathcal{D}_t\partial\widetilde{Y}_s/\partial x_i$, where $\mathcal{D}_t=\mu_t/(\overline{\rho}Sc_t)$ is turbulent diffusivity and $Sc_t=0.5$ is the turbulent Schmidt number. In this paper, Smagorinsky-Lilly model is applied for subgrid scale turbulent eddy-viscosity modelling \citep{smagorinsky1963general}: $\mu_t=\overline{\rho}L_s^2|\widetilde{S}|$, where $L_s$ is the mixing length for subgrid scale and $|\widetilde{S}|\equiv\sqrt{2\widetilde{S}_{ij}\widetilde{S}_{ij}}$. The mixing length $L_s$ is computed using $L_s=C_s\Delta$, where $\Delta$ is local grid scale estimating by $\Delta=V^{1/3}=\Delta x$ and $C_s$ is Smagorinsky constant chosen as 0.1.

After the mathematical model is nondimensionalized, finite volume method is used for discretion. Time marching is dealt with the third-order TVD Runge-Kutta method \cite{gottlieb1998total}. For convection terms, fifth-order WENO scheme \cite{jiang1996efficient} are applied for discretization, while viscous terms are discretized by using the standard central difference method.

\subsection{Initial conditions for OS/JI}
\begin{figure}
   \centering
    \subfigure[Initial conditions]{
    \label{fig: IC parameters} 
    \includegraphics[clip=true,trim=80 80 80 100,width=.85\textwidth]{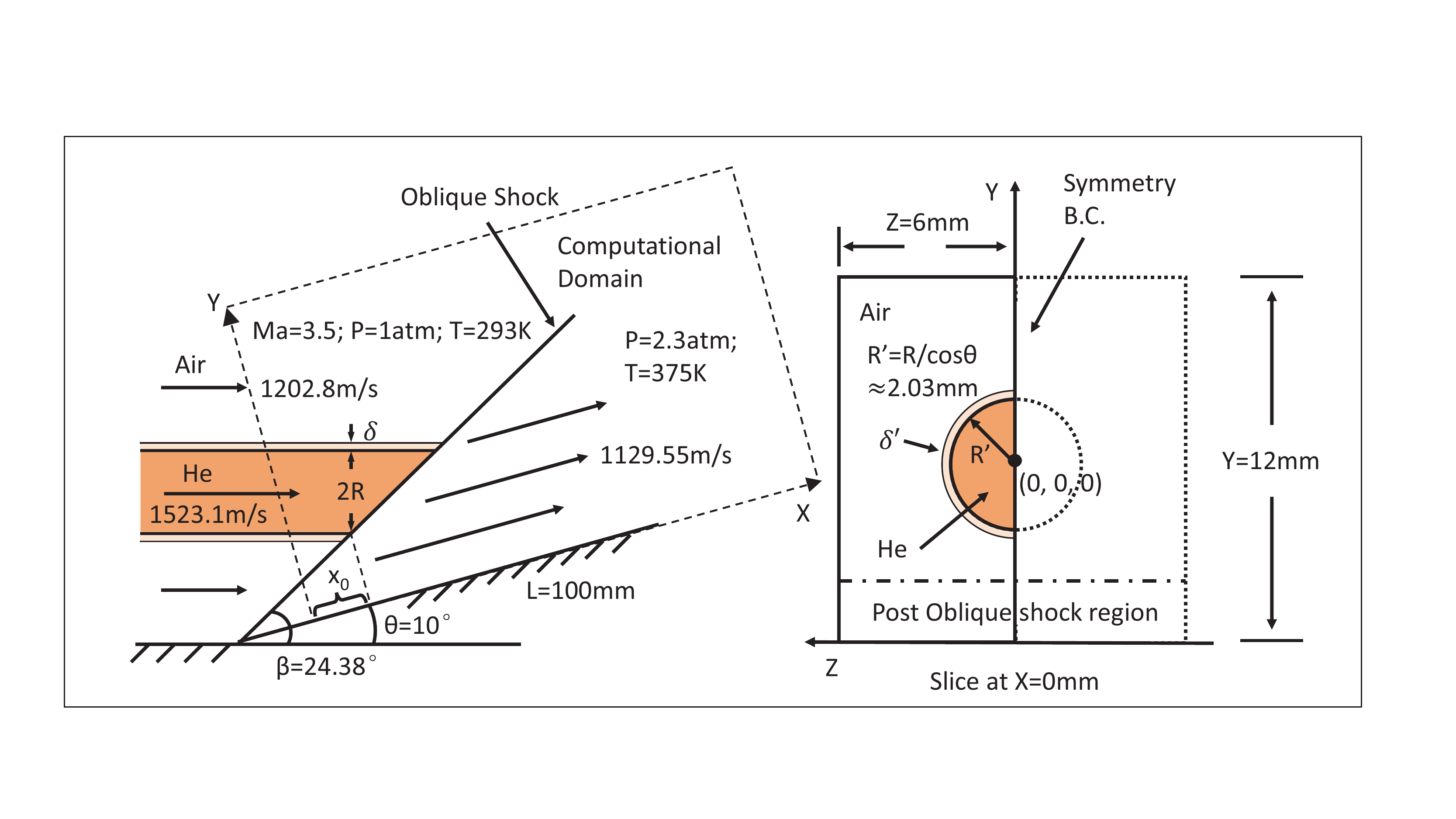}}\\
    \subfigure[Initial conditions for density iso-surface. Different slices along $x$ direction are given. ]{
    \label{fig: IC density} 
    \includegraphics[clip=true,trim=0 0 0 0,width=.7\textwidth]{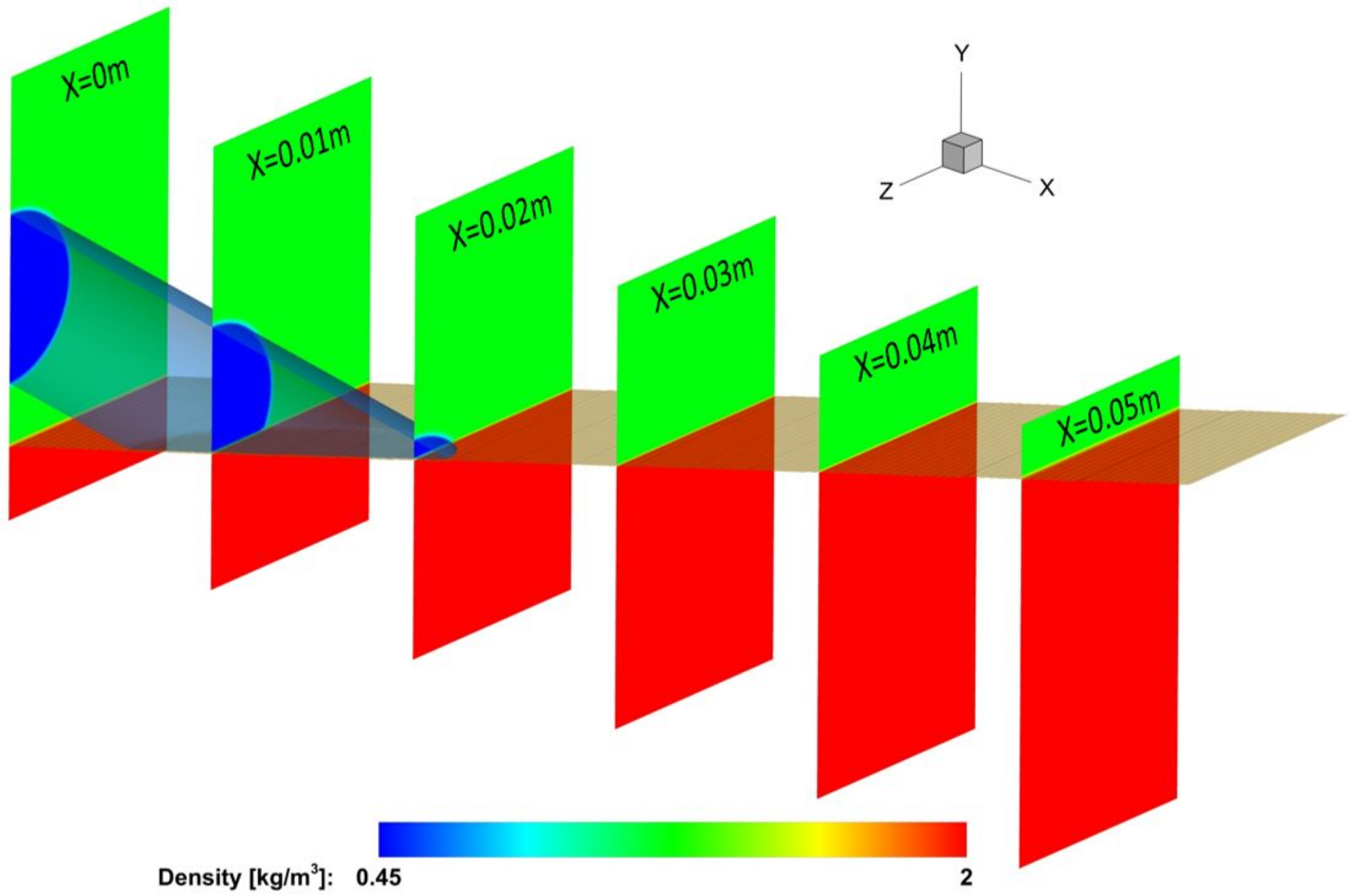}}
    \caption{Initial conditions of oblique shock jet interaction.}\label{fig: initial-conditions}
\end{figure}

The initial condition of OS/JI is illustrated in Fig.\ref{fig: IC parameters}. The coflow of Ma$_{3D}=3.5$ air and Ma$_{3D}=1.5$ Helium jet is set as the incoming gas.
This kind of hypervelocity of internal flow is the typical conditions for sramjet combustor \cite{sislian2006numerical}.
The pressure and temperature is same of both air and Helium jet, which assures that no pressure wave is formed at the outlet of Helium jet.
The radius of the jet is chosen as $R=2mm$, which is similar to the injector diameter of normal injection such as in Hyshot and Hifre~\citep{urzay2018supersonic}.
A diffusive layer is set at the boundary of jet for two reasons.
Firstly, diffusion will naturally occur in the presence of concentration gradient along bubble \cite{tomkins2008experimental}.
Secondly, diffusion layer will avoid the spurious vorticity production along the edge of bubble due to the mesh discretization~\cite{Niederhaus2007}.
In this paper, thickness of diffusion layer is set as $\delta/R=0.2$ with a quasi-Gaussian distribution, which is introduced in our previous studies~\cite{wang2018scaling,li2020gaussian}.

The oblique shock is generated by the deflection of the downside wall. However, in order to simplified the study, the wall condition is not used, but rather a steady oblique shock wave is set at a inclined coordinate whose $x$ direction is chosen as the inclined wall, as shown in left figure of Fig.\ref{fig: initial-conditions}. Although in the inclined coordinate, the slice of Helium jet will become a little bigger because of the inclined projection in the slice of $yz$ plane, deflection of the coordinate is small ($\theta=10^\circ$) that the projection is similar to a circle of radius $R'=2.03mm$. The center of the Helium injection is set as the origin of the coordinate as shown in the right figure of Fig.\ref{fig: IC parameters}. In order to fix the incoming flow condition and the oblique shock wave, the left plane of computational domain is set as the inflow condition and the symmetry boundary is set of the center plane of the bubble. Other boundary conditions are all set as the extrapolation conditions to remove the effect of reflected wave. Fig.\ref{fig: IC density} shows the initial conditions of density distribution in the flow field. An oblique shock cutting the circular Helium jet can be found. The flow after the oblique shock reads the high temperature high pressure status from theoretical oblique shock dynamics, whose parameters can be found in Fig.\ref{fig: IC parameters}.

The grid dependence study is also examined. A locally refinement of mesh is set around the circular jet in order to achieve the high-resolution results and alleviate the numerical computation burden. Three sets of meshes with different resolution have been done. By comparing jet deformation and circulation information, we can infer that the mesh with number of 13.5 million grids has reached grid independence as for the problem studied in present paper. More information can be referred to Appendix.\ref{sec: app1}.

\section{Liff-off phenomena for oblique-shock/Jet interaction}
\label{sec: 3DJSI}
\begin{figure}
  \centering
  \includegraphics[clip=true,trim=0 130 0 0,width=1.0\textwidth]{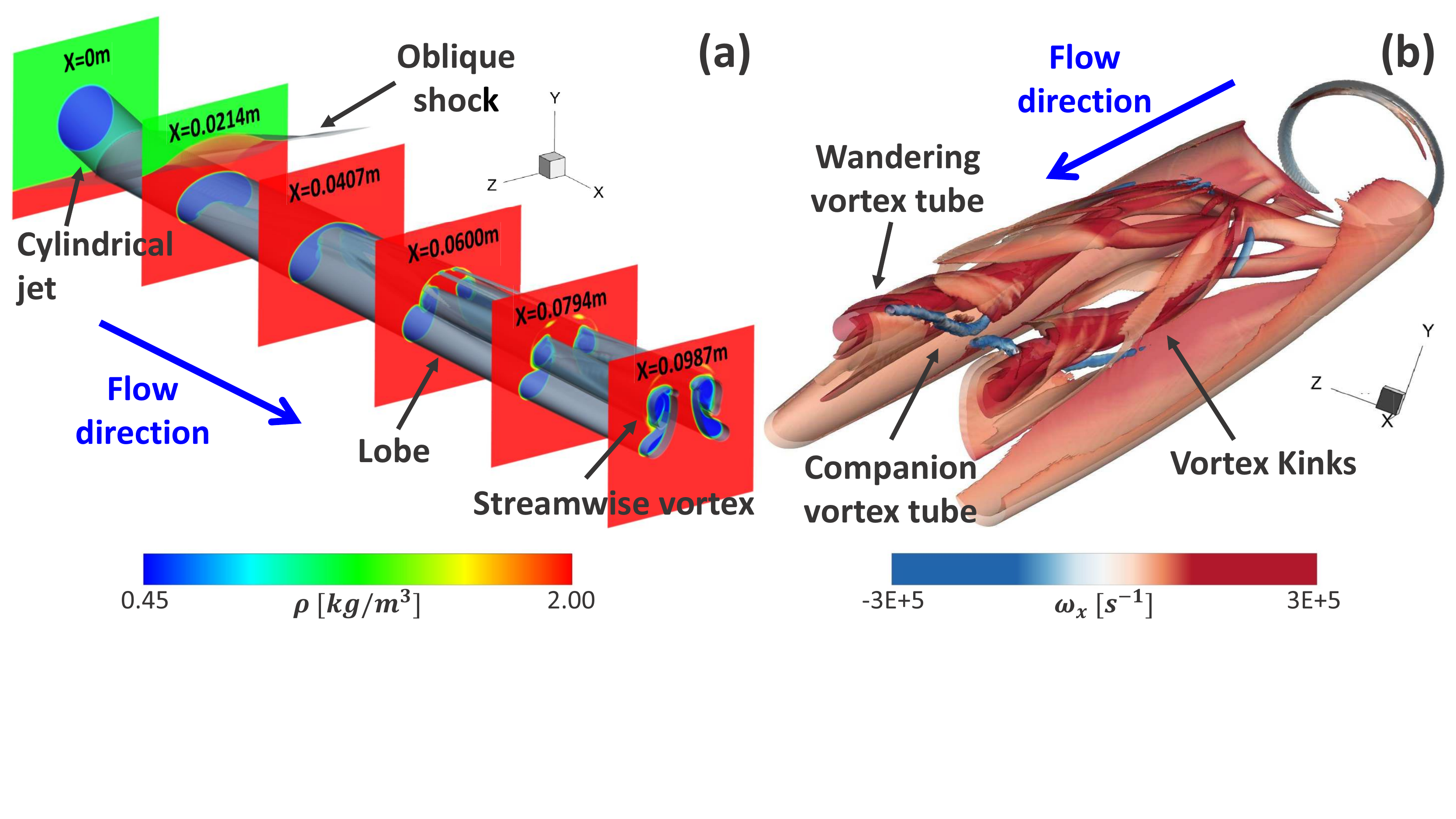}\\
  \caption{(a) Density iso-contour and slices of $yz$ plane along $x$ axis. (b) Iso-contour of $Q$ criterion colored by vorticity magnitude of $x$ direction, $\omega_x$. }\label{fig: 3D OJ/SI}
\end{figure}
Heuristic results of three dimensional oblique shock jet interaction are given in Fig.\ref{fig: 3D OJ/SI}. The results are taken when the flow field has become near steady. Ample fluid phenomena occur in such simple initial geometry.
From density iso-contour as illustrated in Fig.\ref{fig: 3D OJ/SI}(a), a circular jet impinging on a oblique shock is clear. Along the flow direction, the circular jet is deformed into kidney shape as the streamwise vortex emerges. The streamwise vortex benefits from the baroclinic voriticity deposition originating from misalignment of pressure gradient $\nabla p$ of oblique shock and density gradient $\nabla\rho$ from light density jet \cite{ranjan2011shock}. As the streamwise vortex stretches the circular jet along the flow direction, a lobe with high concentration of Helium jet is trailing behind.
The oblique shock is also influenced by the jet penetration as the curvature of the oblique shock shows at the interaction region.
Slices of density contour at $yz$ plane along different position of $x$ axis are inserted into the figure to the density variation along the main flow.
An interesting similarity to two dimensional shock bubble interaction can be found, which will be compared from qualitative and quantitative way in details in the following paper.

In order to study the characteristic of supersonic streamwise vortex, $Q$ criterion \cite{jeong1995identification} colored by vorticity magnitude of $x$ direction, $\omega_x$, was made in Fig.\ref{fig: 3D OJ/SI}(b).
Rigorous vortex tube twisted of the main streamwise vortex wandering along the streamwise direction can be observed. Except from the main streamwise vortex, a companion vortex tube with negative magnitude in vorticity evolves from the very beginning after the oblique shock and rotates closely around the main streamwise vortex. This phenomenon is also called vortex kinks firstly found in the colliding of vortex rings \cite{aref1991linking,chatelain2003reconnection}. We will show later that this companion vortex tube comes from the stretching of the main streamwise vortex.

\begin{figure}
  \centering
  \includegraphics[clip=true,trim=0 140 0 0,width=1.0\textwidth]{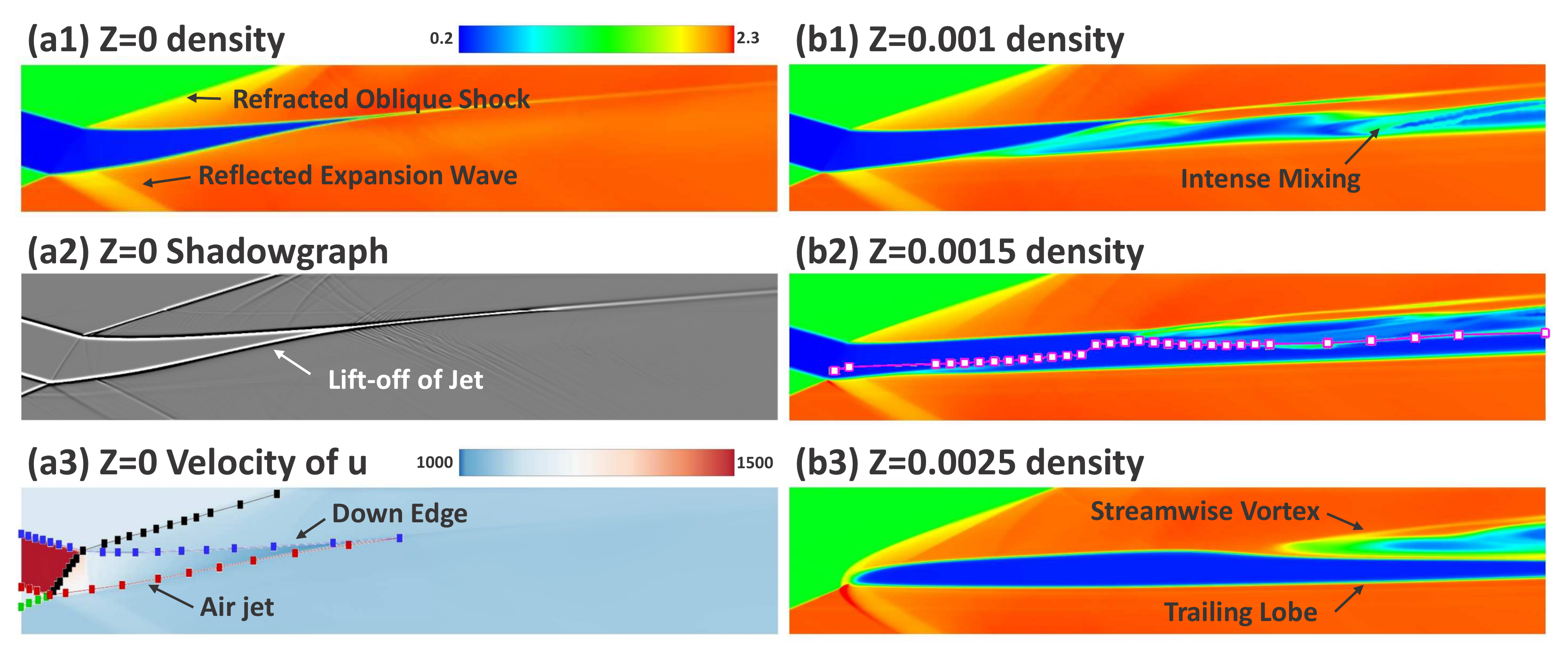}\\
  \caption{(a) Density contour (up), shadowgraph (middle) and velocity of u (bottom) at slice of $z=0$.
                  (b) Density contour of slices at three different position along $z$ axis. }\label{fig: OJ/SI density slice}
\end{figure}
For more details of the OS/JI, slices of $xy$ plane are offered as shown in Fig.\ref{fig: OJ/SI density slice}.
The right part of the Fig.\ref{fig: OJ/SI density slice} shows the parameters of the flow contour (density, shadowgraph, velocity of u) of slice at center slice of $z=0$.
From density contour and shadowgraph, it can find that the oblique shock is refracted inside the Helium jet. A reflected expansion wave is formed at the bottom side of jet.
The penetration of Helium jet at slice of $z=0$ is largely reduced due to oblique shock, which shows mixing induced less Helium concentration downstream from the density contour of Fig.\ref{fig: OJ/SI density slice}(a1).
Lift-off characteristic of the jet is obvious from the structure position of shocked jet as shown in Fig.\ref{fig: OJ/SI density slice}(a3).
The faster Helium jet slows down to a speed that is similar to the post-shock velocity of ambient air from the velocity contour (see Fig.\ref{fig: OJ/SI density slice}(a3)), which suppresses the Kelvin-Helmholz (KH) instability along the streamwise after oblique shock. As for the strong velocity shear before the oblique shock, the KH instability has not been developed due to the short distance before the interaction with oblique shock.

From Fig.\ref{fig: OJ/SI density slice}(b), we shows the the density contour of $xy$ plane at different location along $z$ axis. The intense mixing, presented by the neutral density along the jet, is happening, which is contributed to the strong stretching of supersonic streamwise vortex as shown in Fig.\ref{fig: 3D OJ/SI}. It is noteworthy that at $z=0.0025m$ as shown in Fig.\ref{fig: OJ/SI density slice}(b3), the region where streamwise vortex is not affecting such as lobe region, shows the poor mixing, which emphasizes the importance of mixing enhancement gain from streamwise vortex in supersonic flows.

\section{Nature of lift-off characteristic: an analogy to shock bubble interaction}
\label{sec: 2D3D}
In this section, we will prove that under slender body approximation in frame of oblique shock structure, the lift-off characteristic is inherently and precisely controlled by an analogy to shock bubble interaction.
We first use the spatial-temporal correlation between OS/JI and SBI proposed by Yang \cite{yang1994model}. A general consistency can be observed from the qualitative contour comparison. Moreover, the circulation of streamwise vortex and 2D SBI is compared in quantitative way. However, for the structure position, the correlation fails to link the lift-off structure of OJ/SI and structure kinetics in SBI. Thus, we further propose a objective correlation based on the coordinate of oblique shock, which emerges the striking similarity structures between OS/JI and SBI.

\subsection{Slender-body approximation}
\begin{figure}
   \centering
    \subfigure[Correlation between 2D SBI and 3D OJ/SI.]{
    \label{fig: 2d3d-corr-1} 
    \includegraphics[clip=true,trim=150 90 97.5 75,width=.85\textwidth]{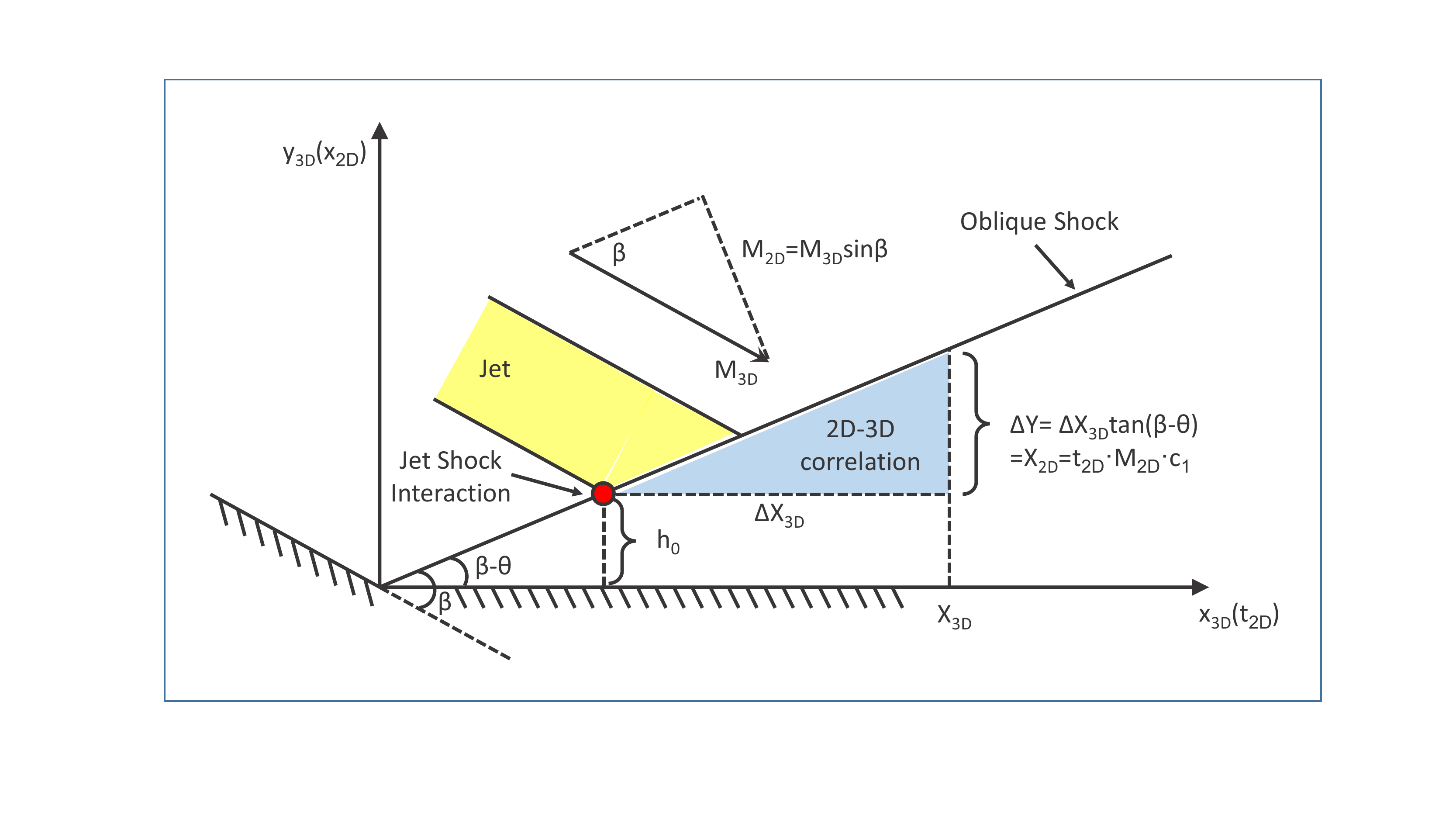}}
    \subfigure[Initial conditions for 2D SBI.]{
    \label{fig: SBI IC} 
    \includegraphics[clip=true,trim=75 120 100 60,width=.85\textwidth]{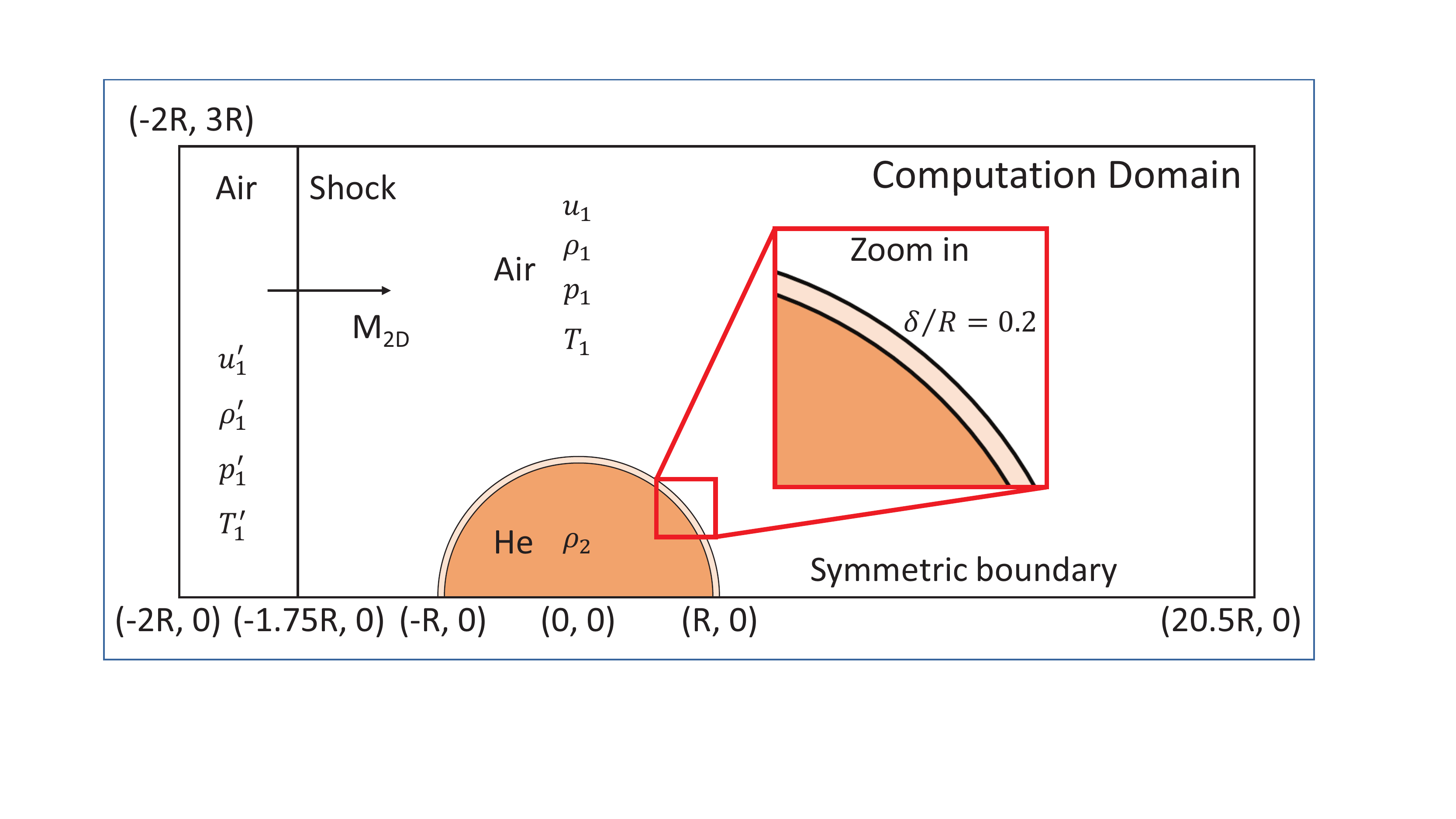}}
    \caption{Correlation for OS/JI and SBI (containing initial conditions for 2D SBI)}\label{fig: 2d3d corr}
\end{figure}
As we can find in Fig.\ref{fig: 3D OJ/SI}, the structures of shocked jet turn to a streamwise vortex with kidney shape that is similar to shock bubble interaction \cite{ranjan2011shock}.
Actually, if the velocity of streamwise flow is larger than the spanwise velocity in one order of magnitude (in this cases $U_{stream}/U_{span}\approx 8$), it is suitable of the slender body approximation, which means the spatial evolution of 3D OS/JI can be analogy to the temporal evolution of 2D SBI \cite{marble1990shock}.
This characteristic enables us to simplify the three-dimensional problem to one of two-dimension when studying the lift-off structure in OS/JI.
Fig.\ref{fig: 2d3d-corr-1} shows an illustration of the core idea of the temporal-spatial correlation between 2D SBI and 3D OS/JI.
Due to the oblique shock wave theory, the actual strength of a oblique shock wave of Ma$_{3D}$ in 3D OS/JI is the same as the normal shock with strength of Ma$_{2D}=$Ma$_{3D}sin\beta\approx1.445$ in 2D SBI.
The key point of correlation is the oblique shock wave is inclined in such way in $x-y$ diagram as the incident normal shock wave of 2D propagates in $t-x$ diagram \cite{yang1994model}:
\begin{equation}\label{eq: sec6-1}
  \Delta Y_{3D}=X_{2D}
\end{equation}
which translates to:
\begin{equation}\label{eq: sec6-2}
  \Delta X_{3D} tan(\beta-\theta)=t_{2D} \cdot Ma_{2D} \cdot c_1
\end{equation}
in which $c_1$ is the sound speed of pre-shock ambient air and $\beta$ can be obtained from the oblique shock wave relation $\beta=\beta($Ma$_{3D},\theta)$ \cite{anderson2010fundamentals}. What needs to be caution is that, as shown in Fig.\ref{fig: 2d3d-corr-1}, distance of $x_0=4mm$ exists before the shock interacts with the Helium jet. Thus, the temporal-spatial correlation considering the length of $x_0$ is represented as:
\begin{equation}\label{eq: 2d3d cor-Yang}
  X_{3D}=x_0+\Delta X_{3D}=x_0+\frac{t_{2D}\cdot Ma_{2D}\cdot c_1}{tan(\beta-\theta)}
\end{equation}

Thus, we further calculated a 2D SBI of initial conditions as shown in Fig.\ref{fig: SBI IC}. The shock Mach number of 2D SBI is set as 1.445, which fixes the post-shock parameters from Rankine-Hugonit relationship \cite{anderson2010fundamentals}. The radius is set same as the circular jet of OS/JI. A diffusion layer of the same distribution is considered as shown in the zoomed insert figure in Fig.\ref{fig: SBI IC}. The computational domain satisfies the temporal evolution of 2D SBI to correlate with 3D OJ/SI from Eq.\ref{eq: 2d3d cor-Yang}. The mesh resolution is chosen the same as the $yz$ plane of 3D OS/JI to restrict the controlling parameters.

\subsection{Qualitative comparison}
\begin{figure}
  \centering
  \includegraphics[clip=true,trim=0 20 0 0,width=1.0\textwidth]{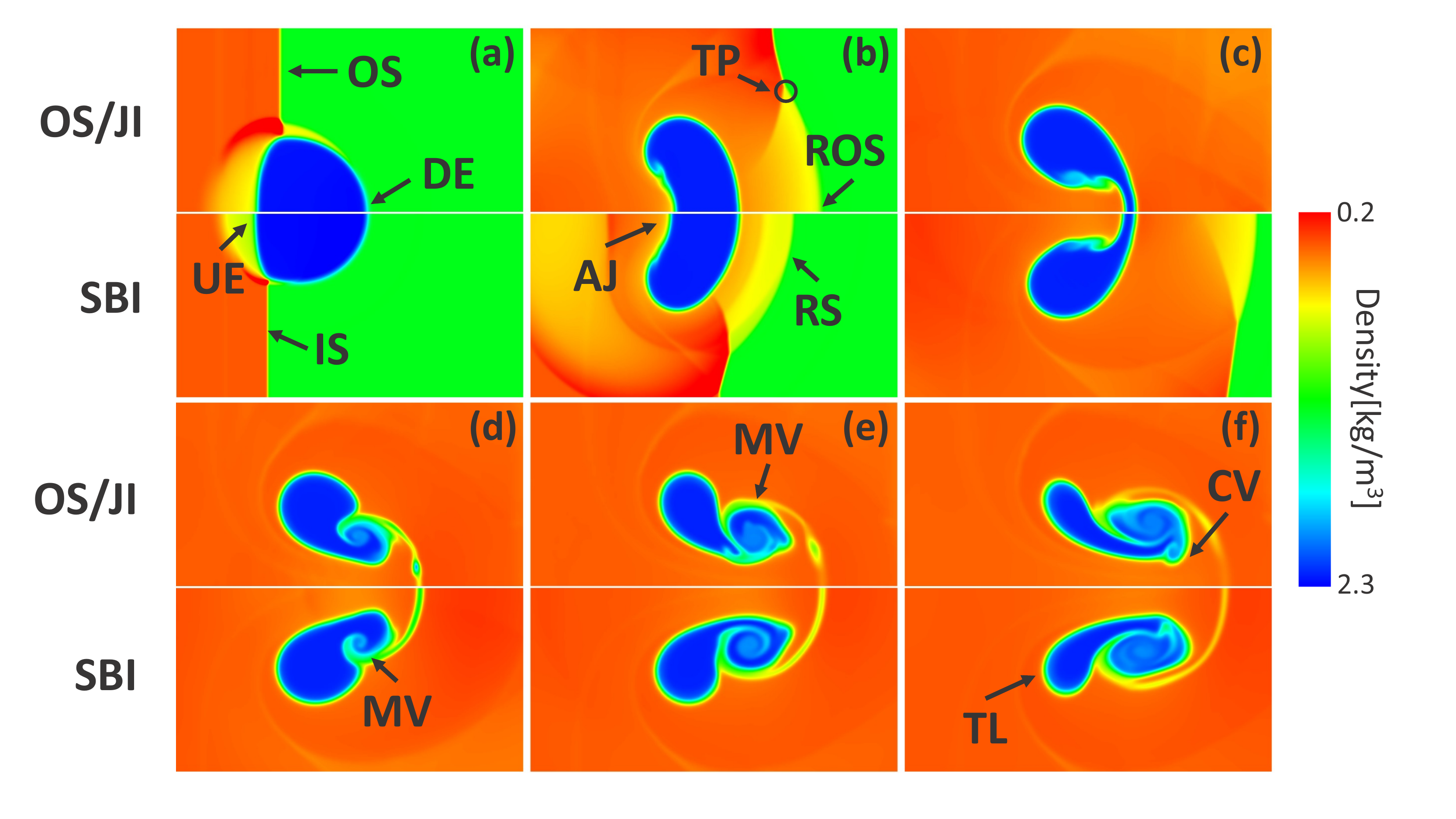}\\
  \caption{Comparison of density contour between 3D OJ/SI and 2D SBI.
  (a) $x=7.87mm$ for 3D and $t=2\mu s$ for 2D SBI, $x/D=1.97$;
  (b) $x=21.4mm$ for 3D and $t=9\mu s$ for 2D SBI, $x/D=5.36$;
  (c) $x=40.8mm$ for 3D and $t=19\mu s$ for 2D SBI, $x/D=10.2$;
  (d) $x=60.2mm$ for 3D and $t=29\mu s$ for 2D SBI, $x/D=15.0$;
  (e) $x=79.5mm$ for 3D and $t=39\mu s$ for 2D SBI, $x/D=19.9$;
  (f) $x=98.9mm$ for 3D and $t=49\mu s$ for 2D SBI, $x/D=24.7$.
  OS: oblique shock; DE: downstream edge; UE: upstream edge; IS: incident shock; AJ: air jet; ROS: refracted oblique shock; RS: refracted shock; MV: main vortex; CV: companion vortex; TL: trailing lobe. }\label{fig: den comp}
\end{figure}
Fig.\ref{fig: den comp} shows the qualitative comparison between 3D OS/JI and 2D SBI. Different slices are extracted from the 3D OS/JI along the streamwise to compare with the temporal evolution of 2D SBI.
The general consistency can be confirmed from the bubble morphology between 2D and 3D results.
After shock impact, the bubble is compressed by shock. The upstream edge moves faster relative to the whole shocked bubble motion, which turns to the air jet (AJ) structure \cite{ranjan2008shock}.  It is this faster motion of AJ that accumulates the centralized baroclinic vorticity to form the main vortex structure, which is also the streamwise vortex structure in 3D OS/JI. From Fig.\ref{fig: den comp}(f), we can observe the formation of companion vortex clearly. It comes from the entrainment process of main vortex during absorbing lobe bubble. This phenomenon also can be confirm by previous experimental results \cite{ranjan2008shock} and is the cause of secondary vortex ring in 3D shock bubble interaction \cite{ranjan2007experimental}.
The exception of structures correlation is the faster refracted shock position in 3D OS/JI than the one in 2D SBI. This violation will be resolved by a change of coordinate, which will be discussed in the following section.

\begin{figure}
  \centering
  \includegraphics[clip=true,trim=0 20 0 0,width=1.0\textwidth]{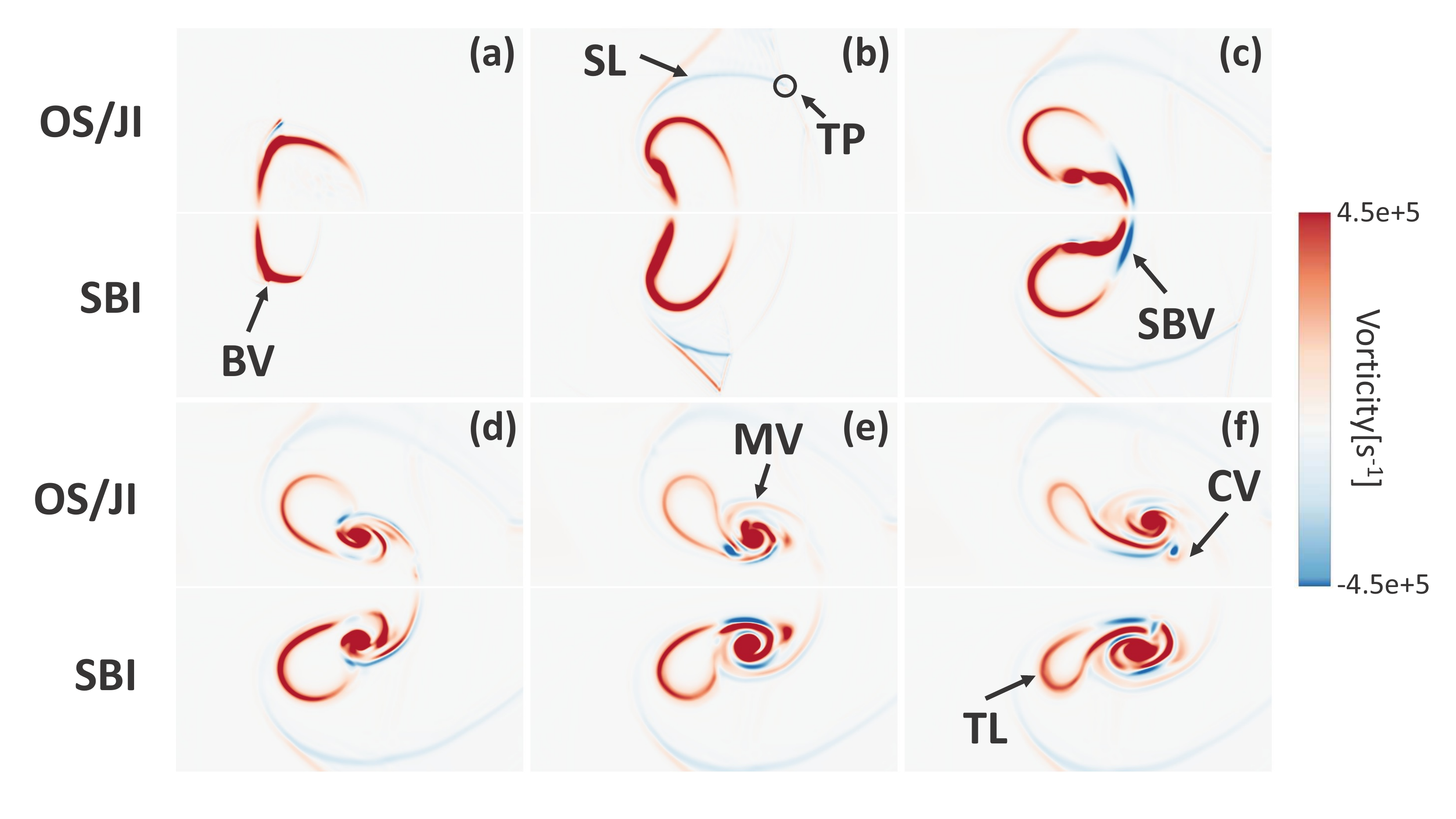}\\
  \caption{Comparison between vorticity contour of $x$ direction in 3D OS/JI and voriticity contour in 2D SBI. Position taken from 3D results and moment taken from 2D results referring to Fig.\ref{fig: den comp}.
  BV: baroclinic vorticity; SL: shear layer; TP: triple point; SBV: secondary baroclinic vorticity.}\label{fig: vor comp}
\end{figure}
The vorticity contour from 3D OS/JI and 2D SBI are also compared in Fig.\ref{fig: vor comp}. The vorticity of $x$ direction in 3D OS/JI is chosen for it takes the dominant value than voritcity magnitude in other two directions.
Baroclinic vorticity is deposited immediately following with the shock passage. The main vortex forms behind air jet structure and gradually absorbs the whole vorticity along the bubble. The secondary baroclinic vorticity production can be found in Fig.\ref{fig: vor comp}(c), which forms into companion vortex in 3D OS/JI. The reason for this secondary baroclinic vorticity production is the local acceleration of light density bubble as revealed in Ref.~\onlinecite{lee2006circulation}. The secondary baroclinic vorticity will shows its effectiveness on accelerating the local mixing rate in RM problem \cite{peng2003vortex}.
Again, the 3D OS/JI and 2D SBI shares the similarities from the qualitative perspective in general.

\subsection{Quantitative comparison}
\begin{figure}
  \centering
  \includegraphics[clip=true,trim=10 25 20 10,width=.85\textwidth]{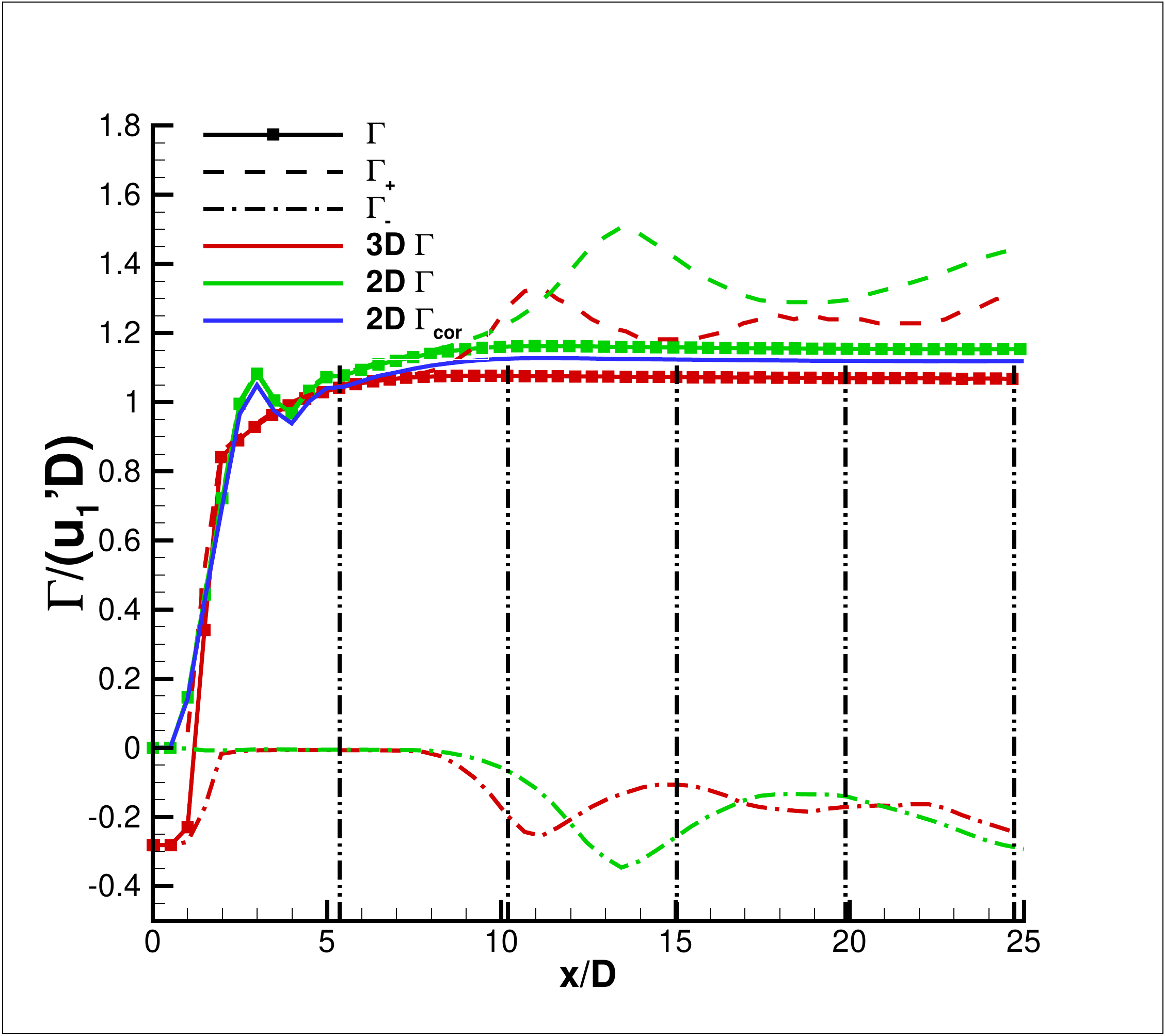}\\
  \caption{Quantitative comparison between circulation of voriticity in $x$ direction $\omega_x$ in 3D OS/JI and in 2D SBI under the spatial-temporal correlation of Eq.\ref{eq: 2d3d cor-Yang}.
  A coordinate corrected circulation of 2D SBI from Eq.\ref{eq: cir_2d-3d} is shown as blue solid line. }\label{fig: circu comp}
\end{figure}
Here, we compare the circulation growth and structure position of 3D OS/JI and 2D SBI in the quantitative way.
First, the circulation of these two cases are calculated as the area integral of vorticity magnitude:
\begin{equation}\label{eq: cirx}
  \Gamma_x=\int\omega_xdA; \quad \Gamma^+_x=\int\omega^+_xdA; \quad \Gamma^-_x=\int\omega^-_xdA
\end{equation}
Moreover, the circulation from secondary baroclinic vorticity is also calculated.
The comparison of circulation is presented in Fig.\ref{fig: circu comp}. Due to the fact that in 3D OJ/SI, the circular jet is injected downwards before interacting with oblique shock (refer to Fig.\ref{fig: initial-conditions}), the circulation is negative at the outlet of circular orifice. It increases immediately after oblique shock interaction, which shows the same trend as circulation growth in 2D SBI. However, the circulation magnitude of 3D OS/JI is slightly lower than one of 2D SBI. Moreover, the secondary baroclinic circulation fails to collapse between 3D case and 2D case.

\begin{figure}
  \centering
  \includegraphics[clip=true,trim=0 0 0 0,width=1.0\textwidth]{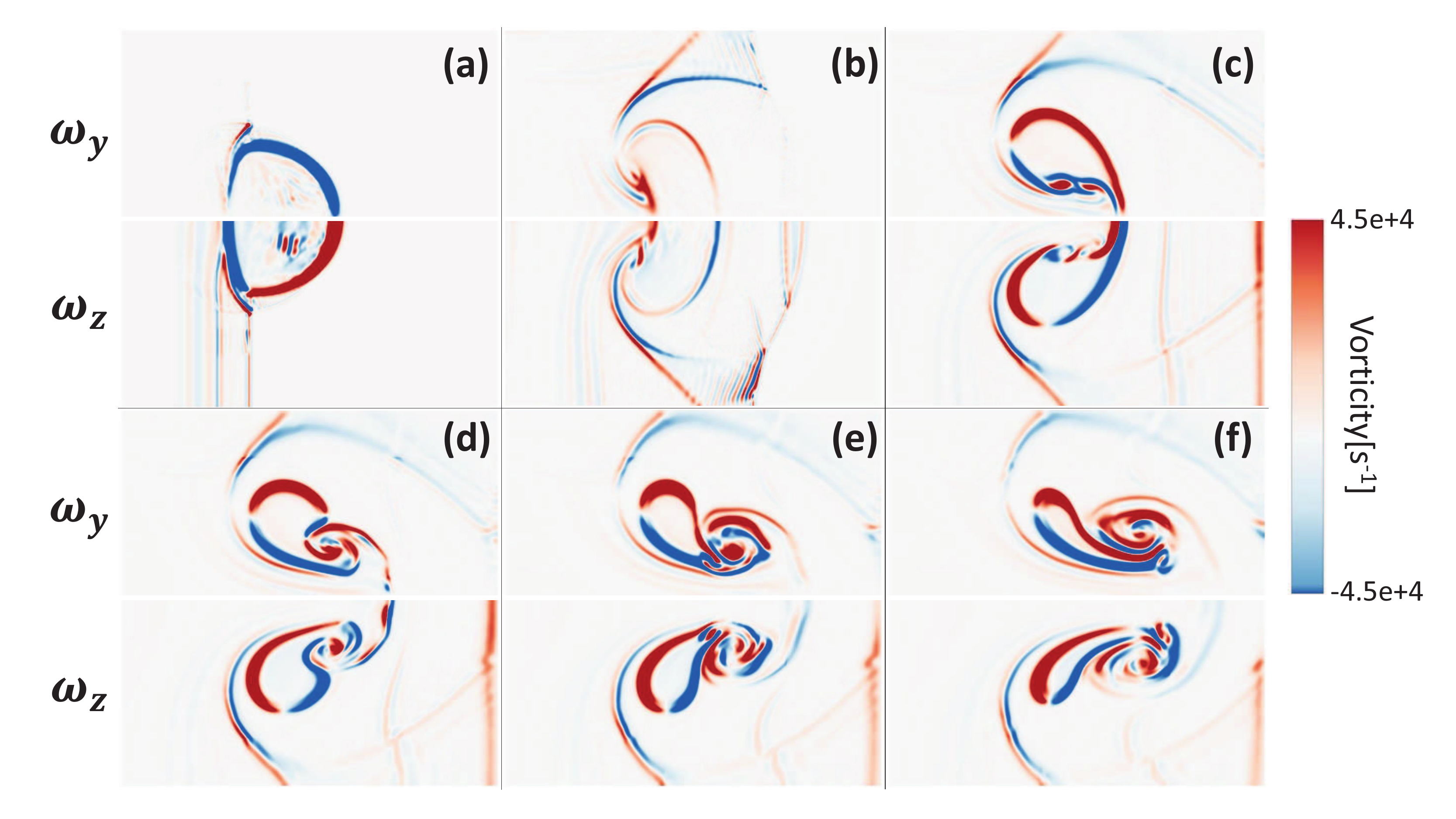}\\
  \caption{Vorticity in $y$ direction and $z$ direction in 3D OS/JI at different positions along $x$ direction (positions taken refer to Fig.\ref{fig: den comp}).}\label{fig: vortex stretching}
\end{figure}
Here, we further seek to the the vorticity evolution in other two direction, namely $\omega_y$ and $\omega_z$ as shown in Fig.\ref{fig: vortex stretching}. It can find that although the vorticity magnitude of this two direction is much lower than one of $x$ direction, considerable value if integrated makes it possible that vorticity in $x$ direction is translated to other direction by vortex stretching \cite{niederhaus2008computational}, which makes the slightly deviation of circulation in 3D OS/JI from the one in 2D SBI.

\begin{figure}
  \centering
  \includegraphics[clip=true,trim=10 10 20 10,width=1.0\textwidth]{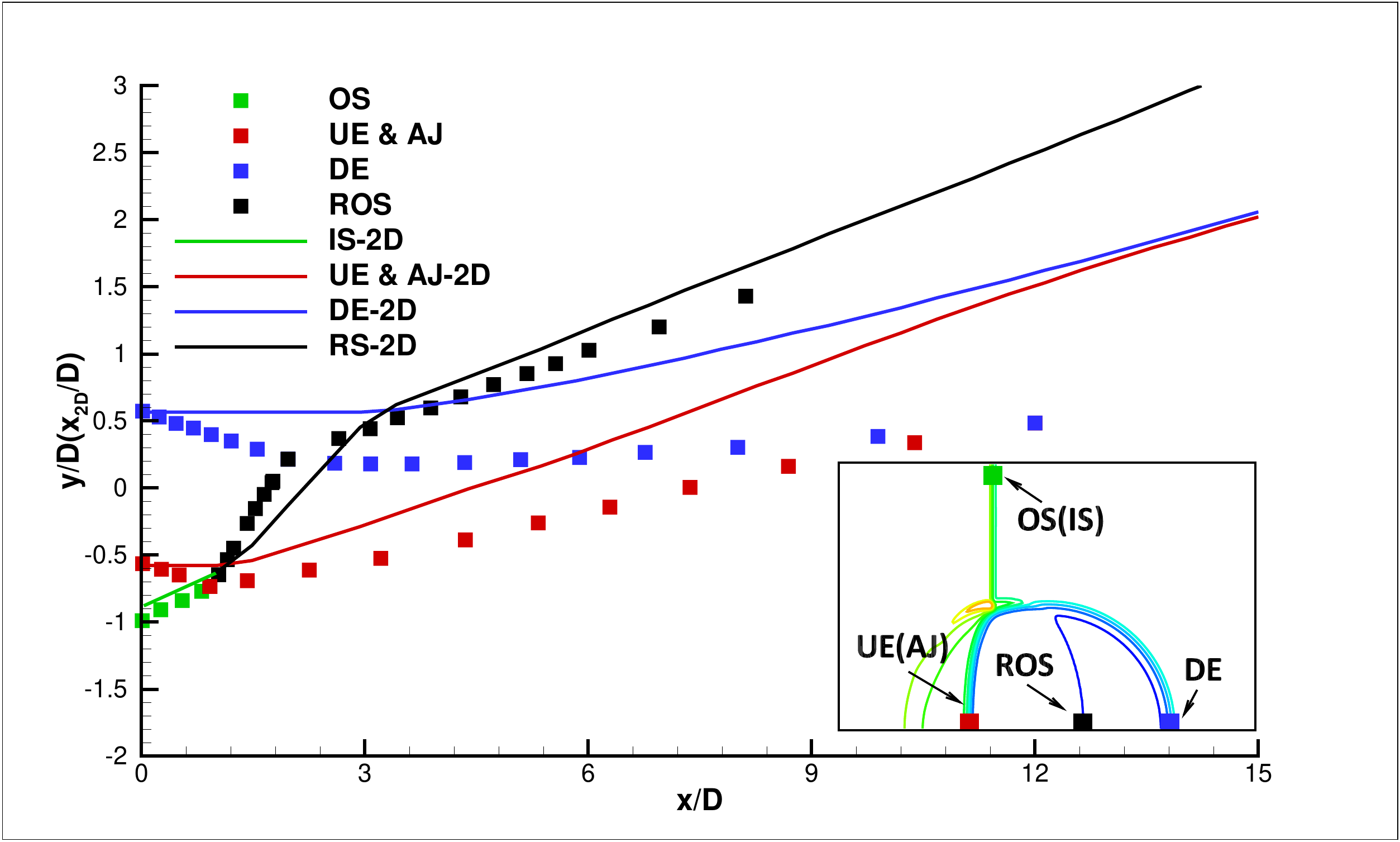}\\
  \caption{Quantitative structure position comparison between 3D OS/JI (obtained from Fig.\ref{fig: OJ/SI density slice}) and 2D SBI under the spatial-temporal correlation of Eq.\ref{eq: 2d3d cor-Yang}. Insert: illustration of structure position. }\label{fig: pos comp}
\end{figure}
The quantitative comparison between structure position in 3D OS/JI and 2D SBI is illustrated in Fig.\ref{fig: pos comp}. The main shock structures and bubble/circular jet structures are faithfully recorded. It interesting to find that the general trend of structure position is similar, while the slopes of structures in 2D SBI are larger than the ones in 3D cases.
However, we find only one structure position is matched, that is the oblique shock in 3D and incident shock in 2D.
This is the key to correct the spatial-temporal correlation of Eq.\ref{eq: 2d3d cor-Yang}.

\subsection{A modified correlation for lift-off characteristics}
\begin{figure}
  \centering
  \includegraphics[clip=true,trim=150 40 150 20,width=1.0\textwidth]{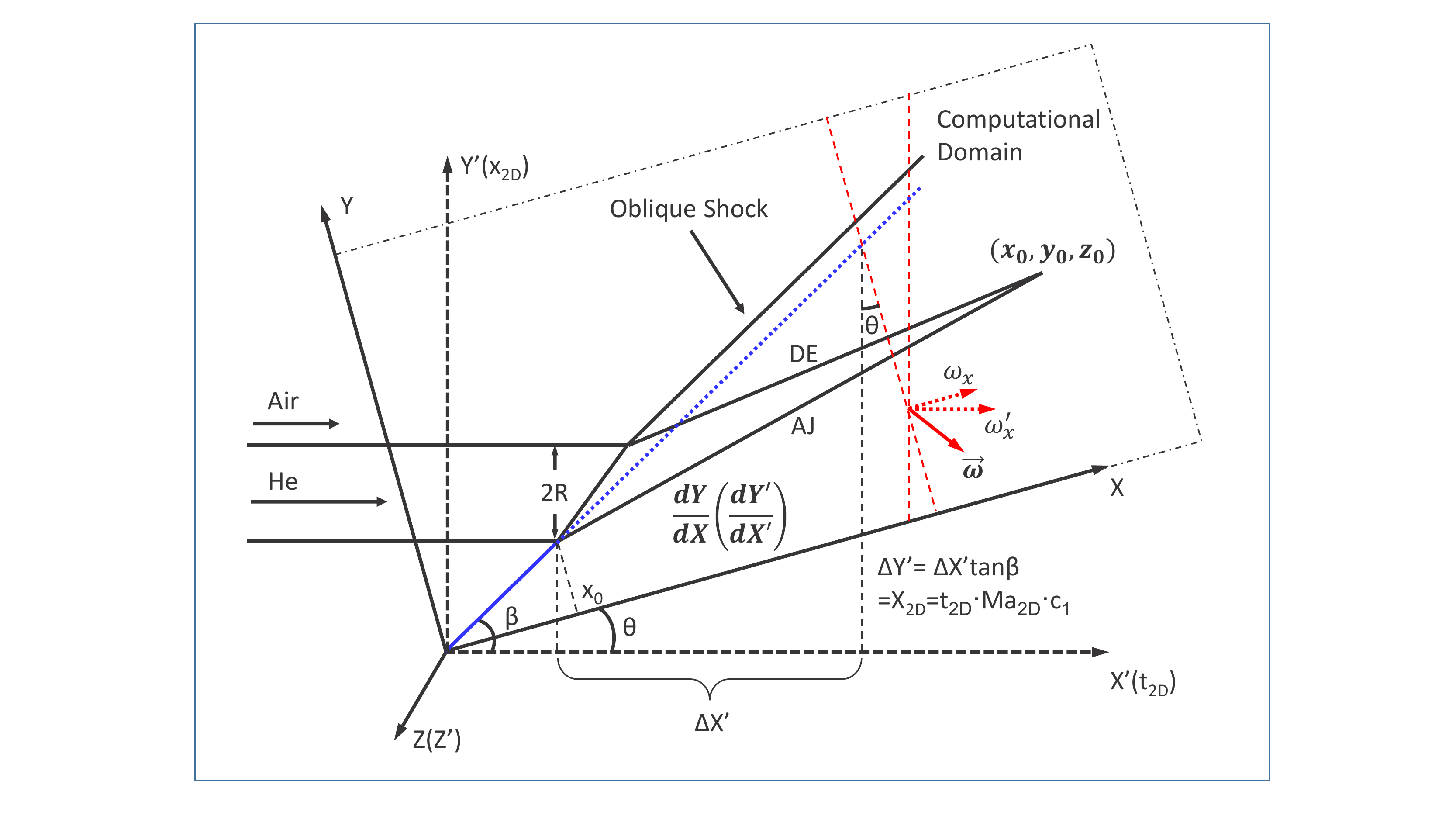}\\
  \caption{An improved spatial-temporal correlation in the objective coordinate frame on oblique shock.}\label{fig: 2d3d corr-new}
\end{figure}
Let us re-examine the initial condition of 3D OS/JI as shown in Fig.\ref{fig: initial-conditions}.
It is naturally to built the correlation of 2D SBI and 3D OS/JI in the frame of computational, which is the laboratory coordinate system convenient to conduct the study \cite{yang1994model}.
However, we may find that temporal-spatial correlation stands only in the coordinate of $Ox'y'z'$ that is on oblique shock wave, which is frame of system with a counter-rotation angle $\theta$ referring to the laboratory system $Oxyz$ as illustrated in Fig.\ref{fig: 2d3d corr-new}.
Then, make correlation between 2D SBI and 3D OS/JI in the coordinate frame of $Ox'y'z'$, we obtain:
\begin{equation}\label{eq: 2d3d cor new}
  X'_{3D}=x'_0+\Delta X'_{3D}=x'_0+\frac{t_{2D}\cdot Ma_{2D}\cot c_1}{tan\beta}
\end{equation}
where $x'_0=x_0cos\beta/cos(\beta-\theta)$ and $Y'=x_{2D}$ needs to be noted.
Thus the position of any structure concerned such as AJ, DE or shock wave structures in new coordinate $Ox'y'z'$ needs to rotate to the coordinate of $Oxyz$ by rotational transformation matrix:
\begin{equation}\label{eq: rot equ}
\left( \begin{array}{c}
\phi_x  \\
\phi_y  \\
\phi_z
\end{array} \right)\quad=\quad
\left( \begin{array}{ccc}
cos\theta & sin\theta & 0  \\
-sin\theta & cos\theta & 0  \\
0 & 0 & 1
\end{array} \right)
\left( \begin{array}{c}
\phi'_x  \\
\phi'_y  \\
\phi'_z
\end{array} \right)
\end{equation}
Thus the position of $(x_0,y_0,z_0)$ can be expressed by:
\begin{equation}\label{eq: 2d3d cor new-2}
\left\{ \begin{array}{l}
 x_0= x'_0cos\theta +y'_0sin\theta  \\
 y_0=-x'_0sin\theta + y'_0cos\theta \\
 z_0=z'_0
\end{array} \right.
\end{equation}

\begin{figure}
  \centering
  \includegraphics[clip=true,trim=10 10 20 10,width=1.0\textwidth]{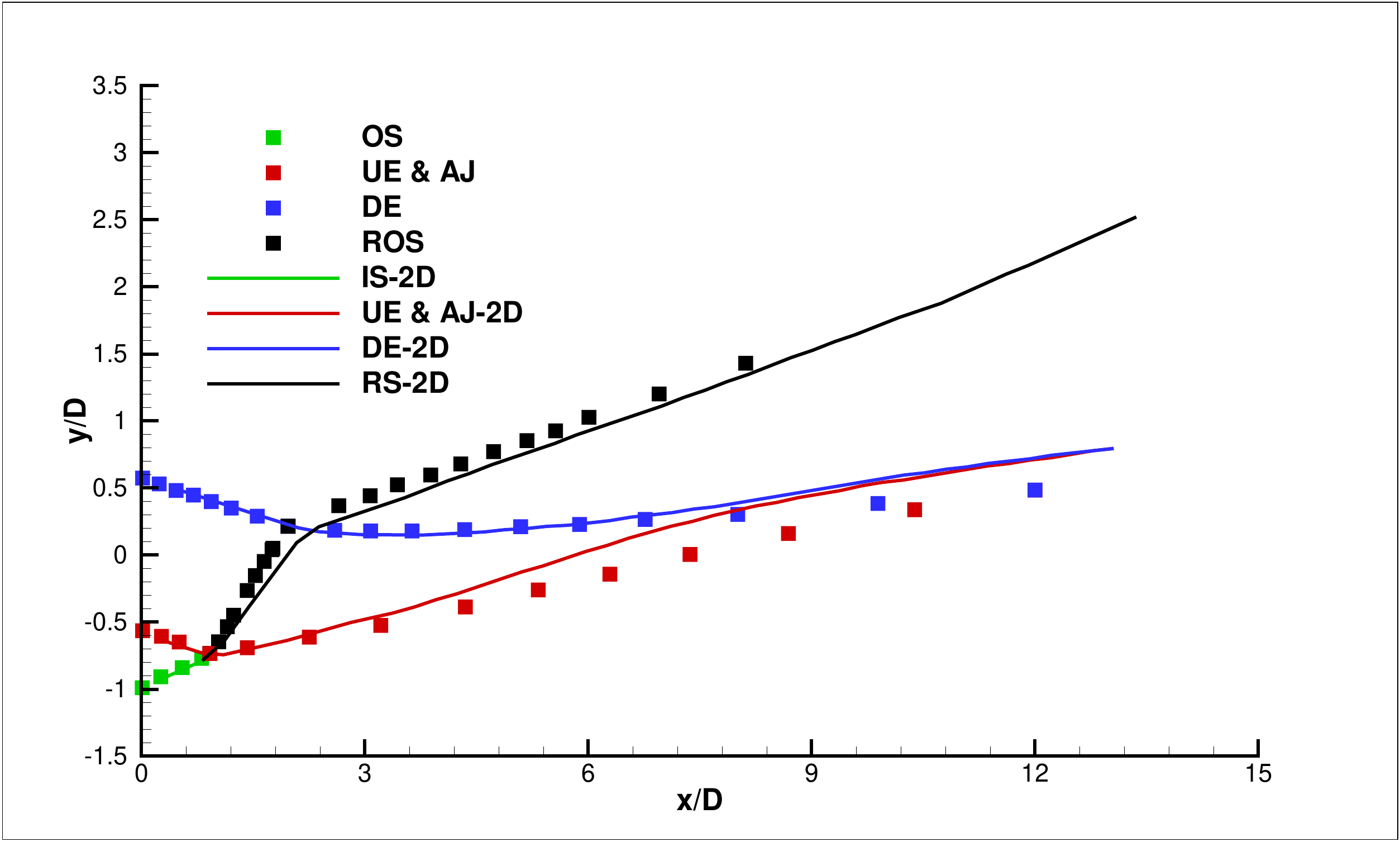}\\
  \caption{Quantitative structure position comparison between 3D OS/JI and 2D SBI under the spatial-temporal correlation of Eq.\ref{eq: 2d3d cor new} (refer to Fig.\ref{fig: pos comp}). The striking similar structures between OS/JI and SBI support the proposition that the lift-off of streamwise vortex is the result of a underlying two-dimensional vortical motion.}\label{fig: pos comp new}
\end{figure}
By change the frame of coordinate concerned from the laboratory system to the objective oblique shock wave system, it can find that position can be correctly matched between 2D SBI and 3D OS/JI as shown in Fig.\ref{fig: pos comp new}.  This shows the intrinsic similarity between 2D and 3D flow structures under the slender body approximation, which will be a useful method the simplified the research in 3D system.

Due to the invariant characteristic of a tensor under different coordinates, any vector should keep the same form so as the curl of velocity vector, which is vorticity vector $\overrightarrow{\textbf{$\omega$}}$:
\begin{equation}\label{eq: vor vec}
  \overrightarrow{\textbf{$\omega$}}=\nabla\times {\textbf{u}}=(\omega_x,\omega_y,\omega_z)=(\omega_{x'},\omega_{y'},\omega_{z'})
\end{equation}
Then the circulation calculated from 2D SBI in coordinate of $Ox'y'z'$ should also also be transformed into the expression in coordinate $Oxyz$:
\begin{equation}\label{eq: cir_2d}
  \Gamma_{x'}=\int\omega_{x'}dA'=\int\omega_{2D}dA_{2D}=\Gamma_{2D}
\end{equation}
In coordinate of $Ox'y'z'$, that assumes $\omega_{y'}=0$ and $\omega_{z'}=0$. This leads to $\omega_x=\omega_{x'}cos\theta$ by rotational transformation of Eq.\ref{eq: rot equ}.
Also considering that the finite volume $dA=dA'cos\theta$, the circulation from 2D SBI in the coordinate of $Oxyz$ should be expressed as:
\begin{equation}\label{eq: cir_2d-3d}
  \Gamma_{x}=\int\omega_{x}dA=\int (\omega_{x'}cos\theta)d(A'cos\theta)=\Gamma_{2D}cos^2\theta.
\end{equation}
The corrected circulation of 2D SBI is also plotted in Fig.\ref{fig: circu comp}. It can find that the correlated circulation of 2D SBI comes near slightly to 3D OS/JI. Yet, the vortex stretching in 3D OS/JI still makes discrepancy as analysed before.

\section{Two-stage growth mode of structure kinetics in shock bubble interaction}
\label{sec: 2D SBI}
In previous section, the correlation between 2D SBI and 3D OS/JI is proposed by introducing the objective coordinate of frame on oblique shock. If the structure lift-off in 3D OS/JI can be directly related to structure velocity in 2D SBI, underlying nature of lift-off can be revealled by dynamics of 2D SBI structure.
This section focuses on building the velocity model of 2D SBI for a series of SBI cases with different shock strength  (from Ma=1.22 to Ma=4).

\subsection{Structure kinetics for a wide range shock strength}
Table.\ref{tab: initial-condi} offers the initial conditions for cases of different shock strength of 2D SBI. The geometry parameters are the same as the ones in previous section shown in Fig.\ref{fig: SBI IC}. Containing shock strength of Ma=1.445, which is compared with 3D OJ/SI case in Sec.\ref{sec: 2D3D}, six cases with different shock Mach number from 1.22 to 4 are simulated to study the velocity discipline of different structures such as air jet, down edge of bubble and vortex center motion.
\begin{table*}
\begin{ruledtabular}
\begin{center}
\begin{tabular}{lllllll}
\multicolumn{1}{c}{Ma} & 1.22     & 1.445    & 1.8      & 2.4      & 3         & 4         \\ \hline
$p_1'$(Pa)                     & 159036.9 & 229891.7 & 366015.0 & 663791.3 & 1046646.5 & 1873802.9 \\
$T_1'$(K)                      & 334.06   & 375.95   & 448.28   & 596.87   & 783.41    & 1182.91   \\
$u_1'$(m/s)                      & 114.71   & 215.75   & 356.58   & 568.31   & 783.41    & 1074.53 \\
$W_i$(m/s)                      & 419.26   & 496.58   & 618.58   &  824.78   & 1030.97    &  1374.62 \\
$u_2'/u_1'$(-)                      & 1.424  & 1.423   & 1.421   & 1.417   & 1.414    & 1.412 \\
$A^+$(-)                      & -0.79  & -0.81   & -0.83   & -0.85   & -0.85    & -0.85
\end{tabular}
\caption{Parameters of different shock Mach number including post-shock pressure $p_1'$, post-shock temperature $T_1'$, post-shock velocity $u'_1$, incident shock velocity $W_i$, $u_2'/u_1'$ and post-shock Atwood number $A^+=(\rho'_2-\rho'_1)/(\rho'_2+\rho'_1)$ respectively calculated from one-dimensional shock dynamics.}
\label{tab: initial-condi}
\end{center}
\end{ruledtabular}
\end{table*}

\begin{figure}
   \centering
    \subfigure[Ma=1.22]{
    \label{fig: 2d velo-1} 
    \includegraphics[clip=true,trim=25 20 50 55,width=.31\textwidth]{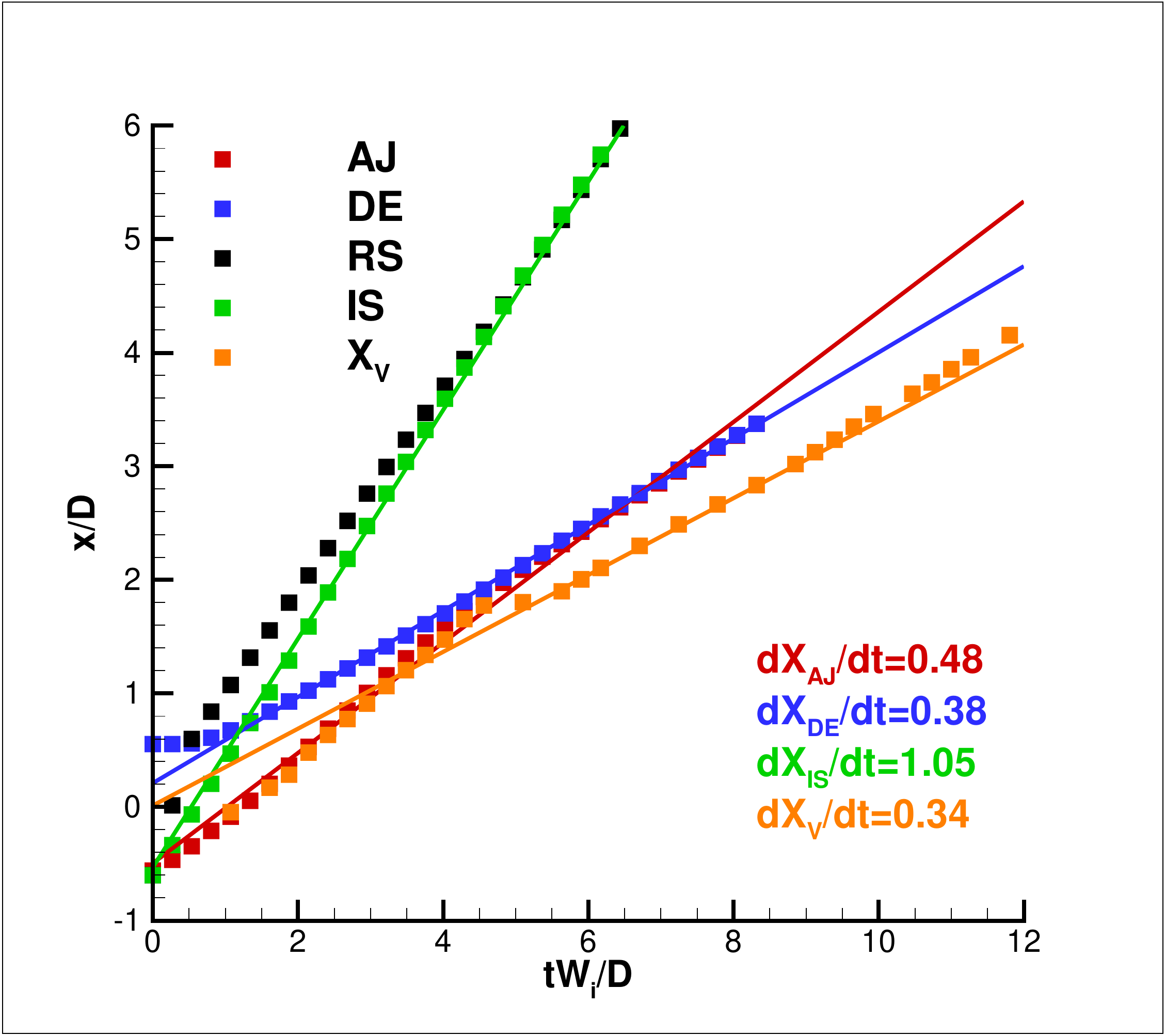}}
    \subfigure[Ma=1.445]{
    \label{fig: 2d velo-2} 
    \includegraphics[clip=true,trim=25 20 50 55,width=.31\textwidth]{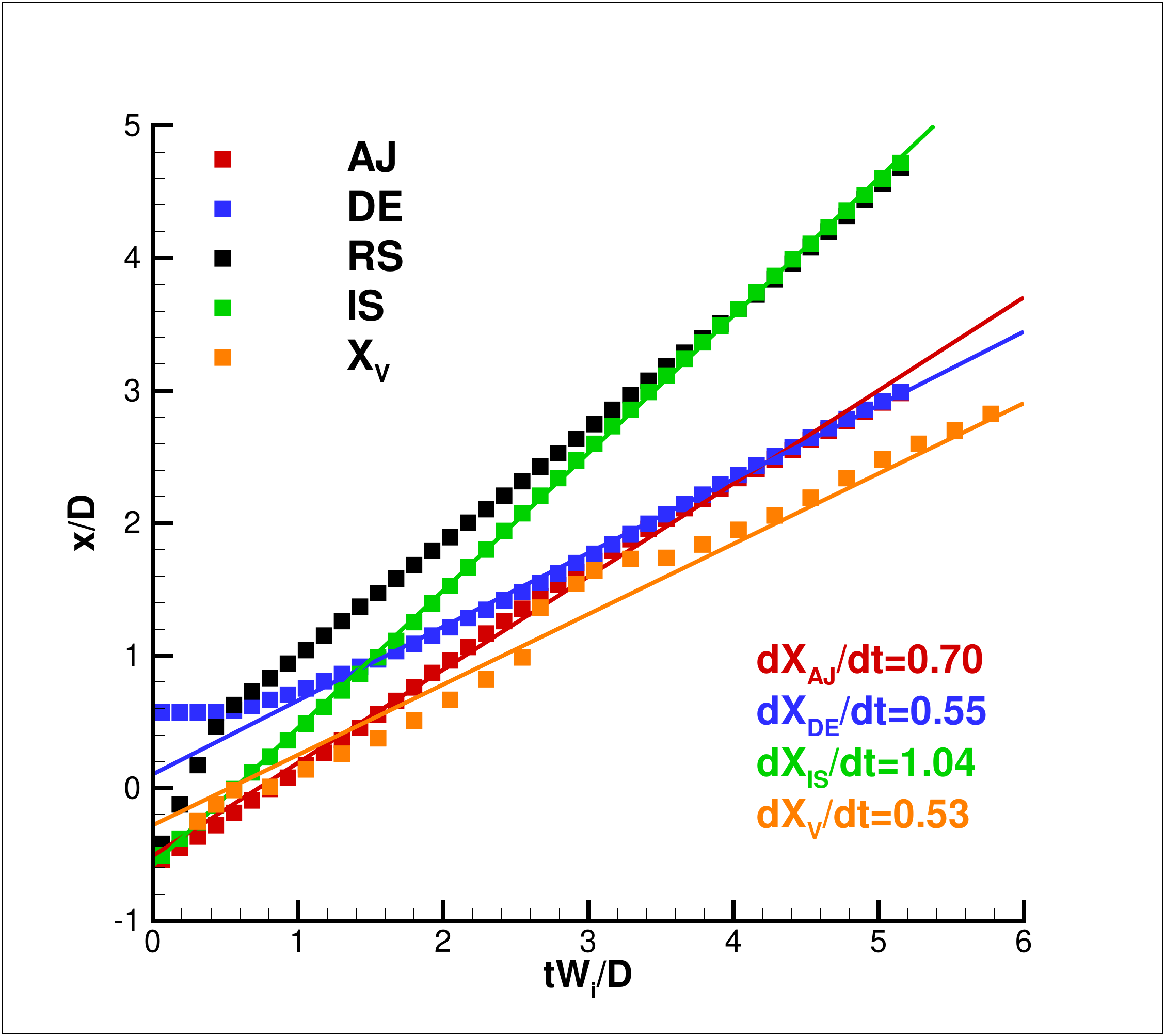}}
    \subfigure[Ma=1.8]{
    \label{fig: 2d velo-3} 
    \includegraphics[clip=true,trim=25 20 50 55,width=.31\textwidth]{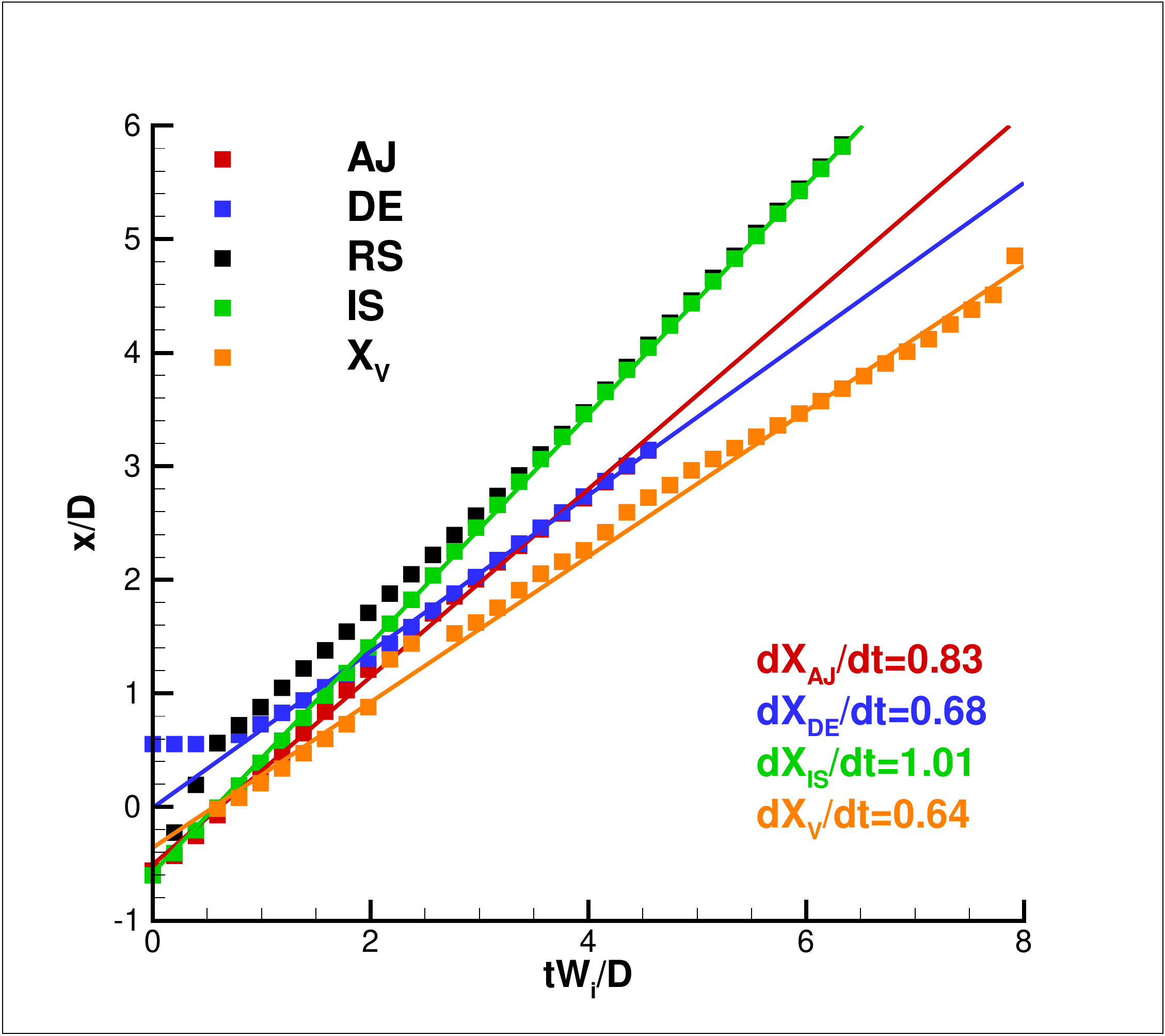}}
    \subfigure[Ma=2.4]{
    \label{fig: 2d velo-4} 
    \includegraphics[clip=true,trim=25 20 50 55,width=.31\textwidth]{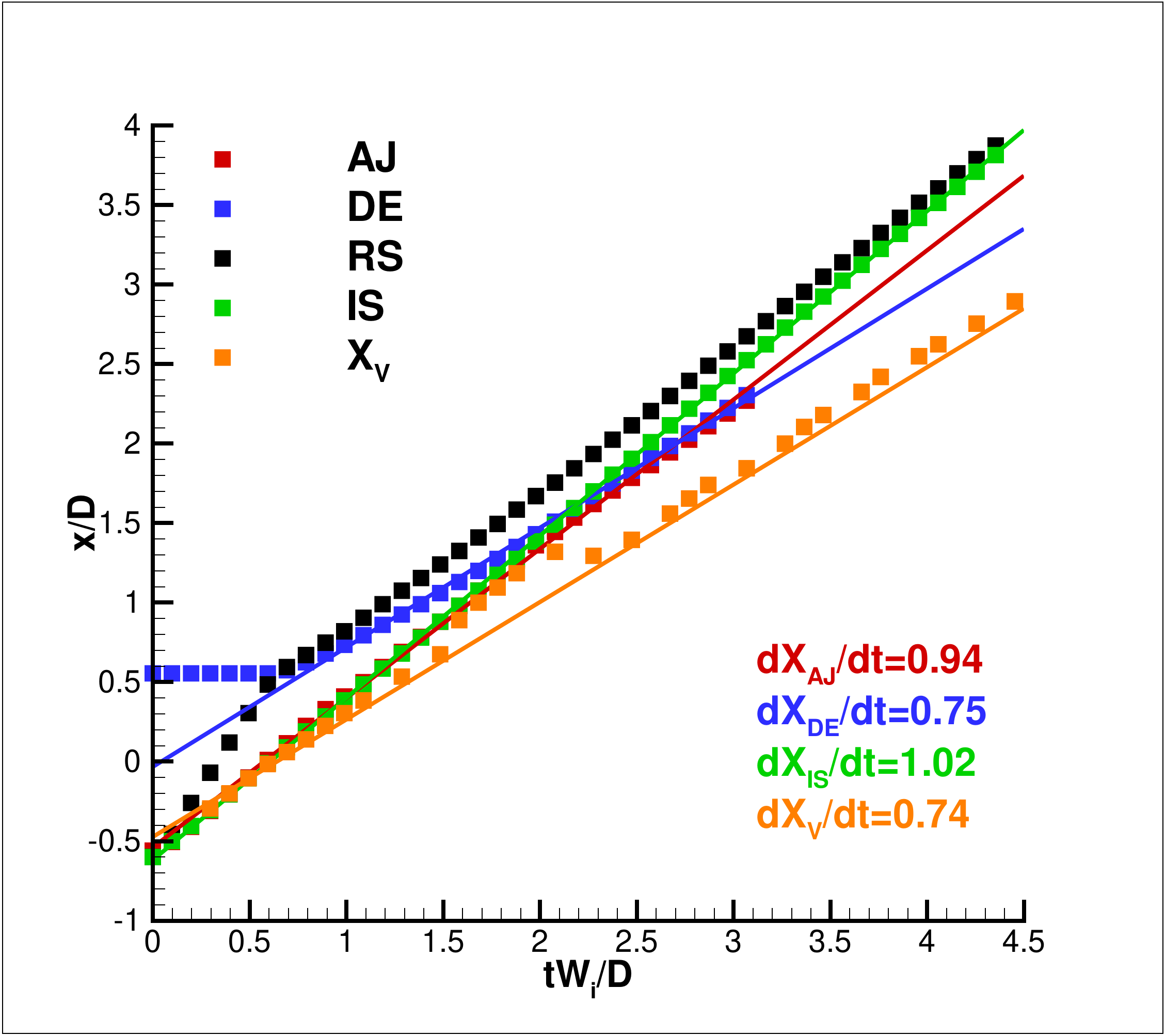}}
    \subfigure[Ma=3]{
    \label{fig: 2d velo-5} 
    \includegraphics[clip=true,trim=25 20 50 55,width=.31\textwidth]{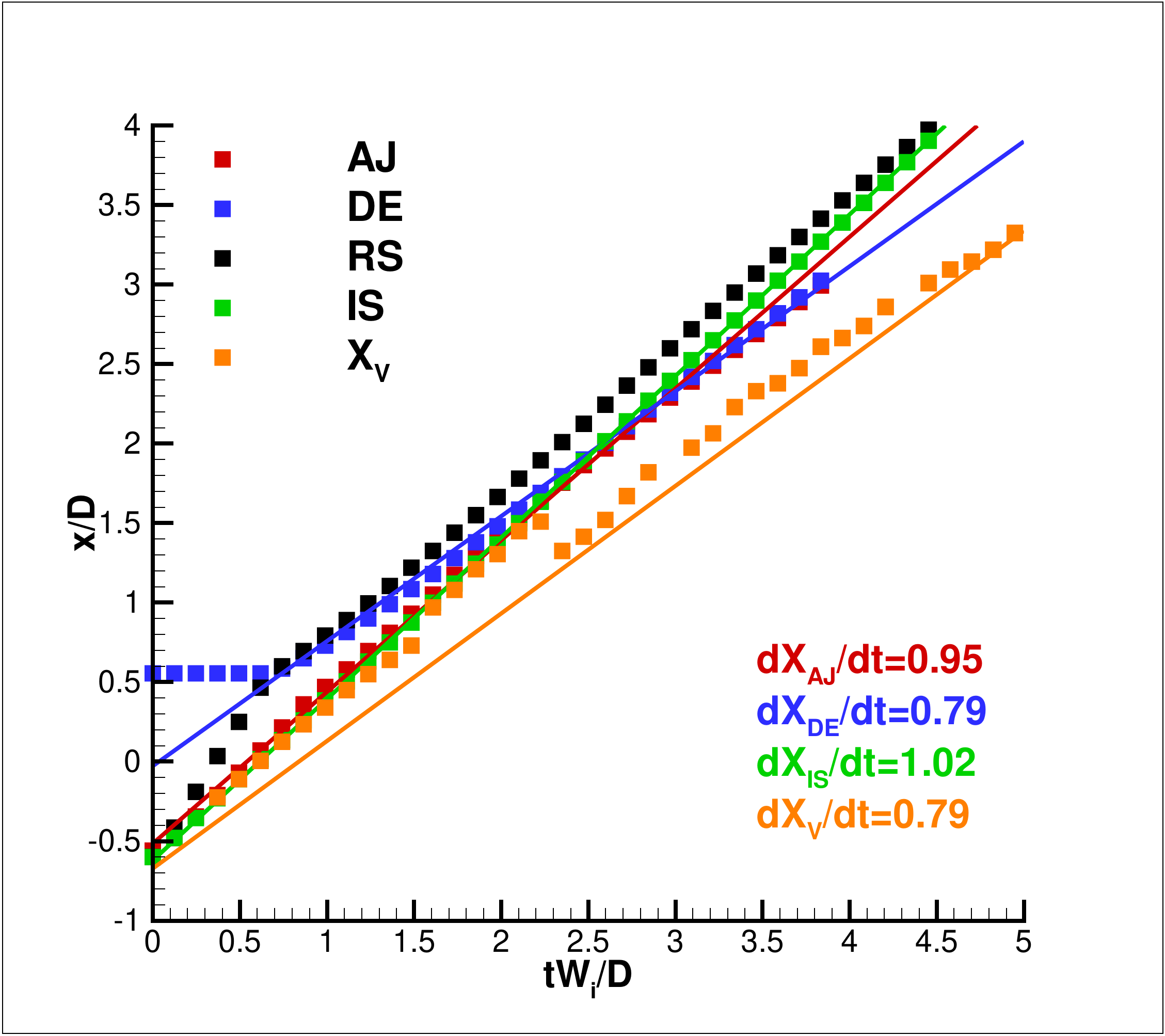}}
    \subfigure[Ma=4]{
    \label{fig: 2d velo-6} 
    \includegraphics[clip=true,trim=25 20 50 55,width=.31\textwidth]{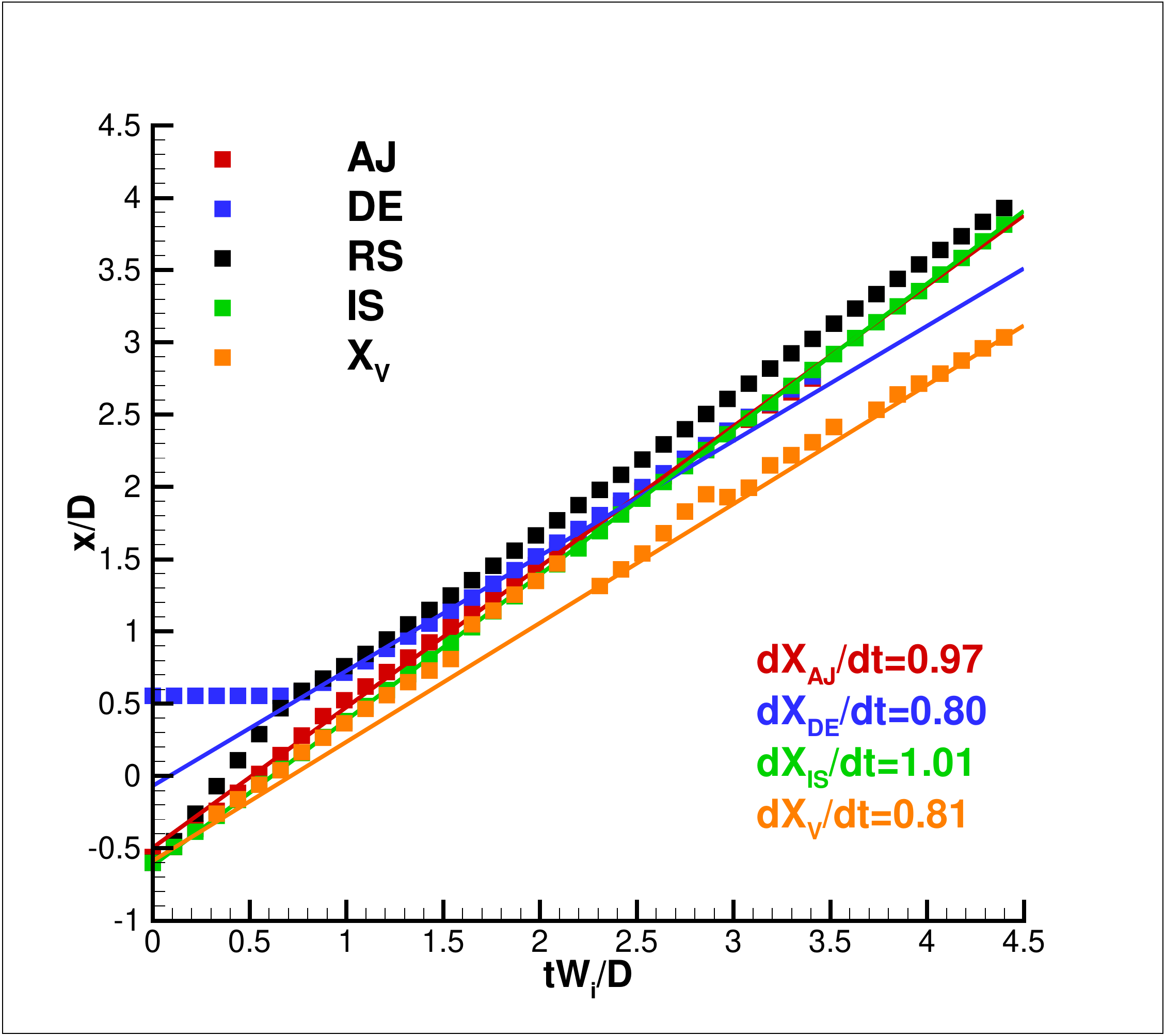}}
    \caption{Structure position in 2D SBI of different Mach number. Linear fit of different structure position are plotted as solid lines whose slopes are remarked in each figure. }\label{fig: 2d velo}
\end{figure}

As shown in Fig.\ref{fig: 2d velo}, positions of different structures of shocked bubble, $x_{2D}/D$, are recorded with variation of dimensionless time $tW_i/D$, where $W_i$ is the incident shock wave speed from one dimension gas dynamics. For all Mach number concerned, speed of incident shock (IS) wave shows general agreement with theoretical values. When refracted shock structure (RS) is imbedded in bubble, faster speed is obtained due the acoustic impedance \cite{ranjan2011shock}. When refracted shock leaves bubble, it converges to incident shock motion at late time. Up stream edge or say air jet (AJ) moves in a faster speed than downstream edge and the speeds of these two structures also become near at late time which forms the bridge structure as shown in Fig.\ref{fig: den comp}. It is interesting to note that the velocity of vortex center is approximately same as the one of down edge of bubble, which means vortex follows downstream edge motion with time and this will be explained in the following model. In general, all structure velocity are higher in stronger shock cases.

\subsection{Review of structure kinetics model in SBI}
In fact, the dynamics of shock bubble interaction has long been a theoretical study focus since the original experiment conducted by Rudinger and Somers \cite{rudinger1960behaviour}. Two theoretical branches of structure kinetics modelling in SBI have emerged. The first is based on the linear growth theory of small perturbation in RMI \cite{richtmyer1960taylor} which is further modified by Haas and Sturtevant \cite{haas1987interaction} (referred to HS model hereafter) through delicate experiment study.
This model \cite{haas1987interaction} treats the SBI as the example of shock-induced Richtmyer-Meshkov instability \cite{brouillette2002richtmyer}. The core idea is shown in the left part of Fig.\ref{fig: illu of modi velo model}(a). As the analogy, SBI is regarded as the perturbation with curvature in amplitude $a$ and wave number $k$ much larger than small sinusoidal interface widely-accepted in single-mode RMI problems \cite{Jacobs2005Experiments}.
Then the upstream of bubble or say air jet point can be regarded as the crests and downstream edge of bubble can be regarded as troughs in the model of linear growth in RMI. The first validation of HS model is numerical simulation carried by Picone and Boris \cite{picone1988vorticity} and is further modified by Ding~\emph{et al}~\cite{Ding2017On} through introducing a reduction factor $\phi$  for air jet structure~\cite{Rikanati2003High}:
\begin{equation}\label{eq: AJ}
  \frac{u_{\textrm{AJ}}}{u'_2}=1-\phi Z_cakA^{+}
\end{equation}
where $u_2'$ is the velocity jump of upstream interface from shock impact. $Z_c=1-u_2'/W_i$ is the shock induced compression factor and $A^+$ is the post-shock Atwood number. For amplitude $a$ and wave number $k$, it maintains $ak=1$ in the cylinder case \cite{haas1987interaction}. The reduction factor $\phi$ is suggested as 0.6 in light density bubble~\cite{Ding2017On} and as 0.5 in heavy density bubble \cite{ding2018interaction}.
Then for the down edge of bubble, the velocity is sightly smaller than the velocity jump of shocked gas $u'_2$:
\begin{equation}\label{eq: DE}
  \frac{u_{\textrm{DE}}}{u'_2}=1+\phi Z_cakA^{+}
\end{equation}

\begin{table*}
\begin{ruledtabular}
\begin{center}
\begin{tabular}{lllllll}
\multicolumn{1}{c}{Ma} & AJ[Eq.\ref{eq: AJ}]   & AJ{[}meas{]} & $\varepsilon(\%)$  & DE[Eq.\ref{eq: DE}]   & DE{[}meas{]} & $\varepsilon(\%)$    \\ \hline
1.22                   & 0.50 & 0.48         & 3.73 & 0.28 & 0.38         & -27.01 \\
1.445                  & 0.73 & 0.70         & 4.31 & 0.50 & 0.56         & -9.62  \\
1.8                    & 0.89 & 0.83         & 7.99 & 0.75 & 0.69         & 8.35   \\
2.4                    & 0.99 & 0.94         & 5.24 & 0.96 & 0.75         & 28.30  \\
3                      & 1.02 & 0.95         & 7.17 & 1.07 & 0.79         & 36.68  \\
4                      & 1.05 & 0.97         & 7.56 & 1.16 & 0.80         & 46.20
\end{tabular}
\caption{Comparison between dimensionless velocity obtained from theoretical model (HS model \cite{haas1987interaction}) and numerical result (measured from Fig.\ref{fig: 2d velo}) for air jet structure and down edge structure. Error of theoretical value is also given.}
\label{tab: HS model}
\end{center}
\end{ruledtabular}
\end{table*}
Table.\ref{tab: HS model} compares the structure velocity of AJ and DE obtain from theoretical model and numerical results in Fig.\ref{fig: 2d velo}. It can find that HS model lightly over-predicts the results within an acceptable error range ($\varepsilon<8\%$), which coincides with the results reported in Ref.~\onlinecite{haas1987interaction}.
However, for downstream edge, large violation of theoretical value is observed from numerical results.
It is noteworthy that this phenomenon is also recorded in Ref.~\onlinecite{haas1987interaction} where general smaller value of experimental results than the prediction from theoretical value. This shows that as far as downstream edge is concerned, the linear RM type HS model will not be suitable in SBI problem. The reason will be analyzed below and a modified downstream edge prediction model is proposed accordingly.

Due to the fact that curvature of bubble is normally much larger than small perturbation, which makes the HS model doubtful for prediction of nonlinear vortex dynamics for late-time behavior.
Thus, another branch is based on the theory of the propagation of a vortex ring under the impulsive acceleration of a disk proposed by Taylor \cite{taylor1953formation}. Focusing on the late time behavior of vortex motion, a vortex propagation model for SBI is proposed by Rudinger and Somers \cite{rudinger1960behaviour} (referred to RS model hereafter).
In view of the vortex dynamics, RS model supposes the bubble as the solid particle which will be accelerated by shock in the speed of $u_p$ by invoking the well-known concept of `apparent additional mass' (virtual mass):
\begin{equation}\label{eq: AJ-2}
  \frac{    u_{\textrm{P}}}{u'_1}=\frac{1+k}{\sigma+k} \qquad \textrm{where }  \sigma=\frac{\rho'_2}{\rho'_1}
\end{equation}
and additional mass fraction $k=0.5$ for spherical bubble and $k=1.0$ for cylindrical bubble ($k$ is also called inertial coefficient).
Thus, velocity of vortex motion can be modelled through the vortex formation from a sudden start of a disk as:
\begin{equation}\label{eq: vor}
  \frac{u_{\textrm{V}}}{u'_1}=1+\frac{2}{\pi^2}\frac{1-\sigma}{\sigma+1}
\end{equation}
In fact, we can find that in RS model, vortex motion is actually formed from the motion of air jet~\cite{haas1987interaction}.
As analyzed in Fig.\ref{fig: vor comp}, the vortex core become evident once air jet is penetrates into the bubble, which is illustrated in Fig.\ref{fig: illu of modi velo model}(d).
Moreover, in Ref.~\onlinecite{Layes2009Experimental}, bubble motion is also treated as air jet, which is  $u_{\textrm{P}}=u_{\textrm{AJ}}$.

While vortex propagation model has been widely used \cite{jacobs1992shock,quirk1996dynamics,ranjan2008shock,zou2016aspect}, it is noteworthy that the discrepancy between theoretical value and actual value obtained either from experiment or from numerical simulation is quite large in most cases \cite{Layes2009Experimental,mori2012suppression}.
Here, we offer Table.\ref{tab: RS model} that compares the structure velocity of AJ and vortex motion obtained from theoretical model and numerical results in Fig.\ref{fig: 2d velo}.
It can find that although RS model predicts the vortex velocity at shock Mach number lower than 1.8, it over-predicts the vortex speed especially in high shock Mach number. Also, AJ structure is hardly captured by RS model.
\begin{table*}
\begin{ruledtabular}
\begin{center}
\begin{tabular}{lllllll}
\multicolumn{1}{c}{Ma} & $u_p$[Eq.\ref{eq: AJ-2}] & AJ{[}meas{]} & $\varepsilon(\%)$   & Vortex[Eq.\ref{eq: vor}]  & Vortex{[}meas{]} & $\varepsilon(\%)$   \\ \hline
1.22                   & 0.49 & 0.48         & 1.15  & 0.32 & 0.34          & -6.48 \\
1.445                  & 0.79 & 0.70         & 11.96 & 0.51 & 0.53          & -4.74 \\
1.8                    & 1.05 & 0.83         & 27.60 & 0.67 & 0.64          & 5.24  \\
2.4                    & 1.27 & 0.94         & 35.56 & 0.81 & 0.73          & 9.93  \\
3                      & 1.37 & 0.95         & 43.72 & 0.87 & 0.79          & 9.81  \\
4                      & 1.45 & 0.97         & 48.87 & 0.92 & 0.82          & 12.18
\end{tabular}
\caption{Comparison between dimensionless velocity obtained from theoretical model (RS model \cite{rudinger1960behaviour}) and numerical result (measured from Fig.\ref{fig: 2d velo}) for equivalent piston structure (comparing with air jet) and vortex structure. Error of theoretical value is also given.}
\label{tab: RS model}
\end{center}
\end{ruledtabular}
\end{table*}

\subsection{Structure kinetics based on two-stage growth mode}
\begin{figure}
  \centering
  \includegraphics[clip=true,trim=0 125 0 0,width=1.0\textwidth]{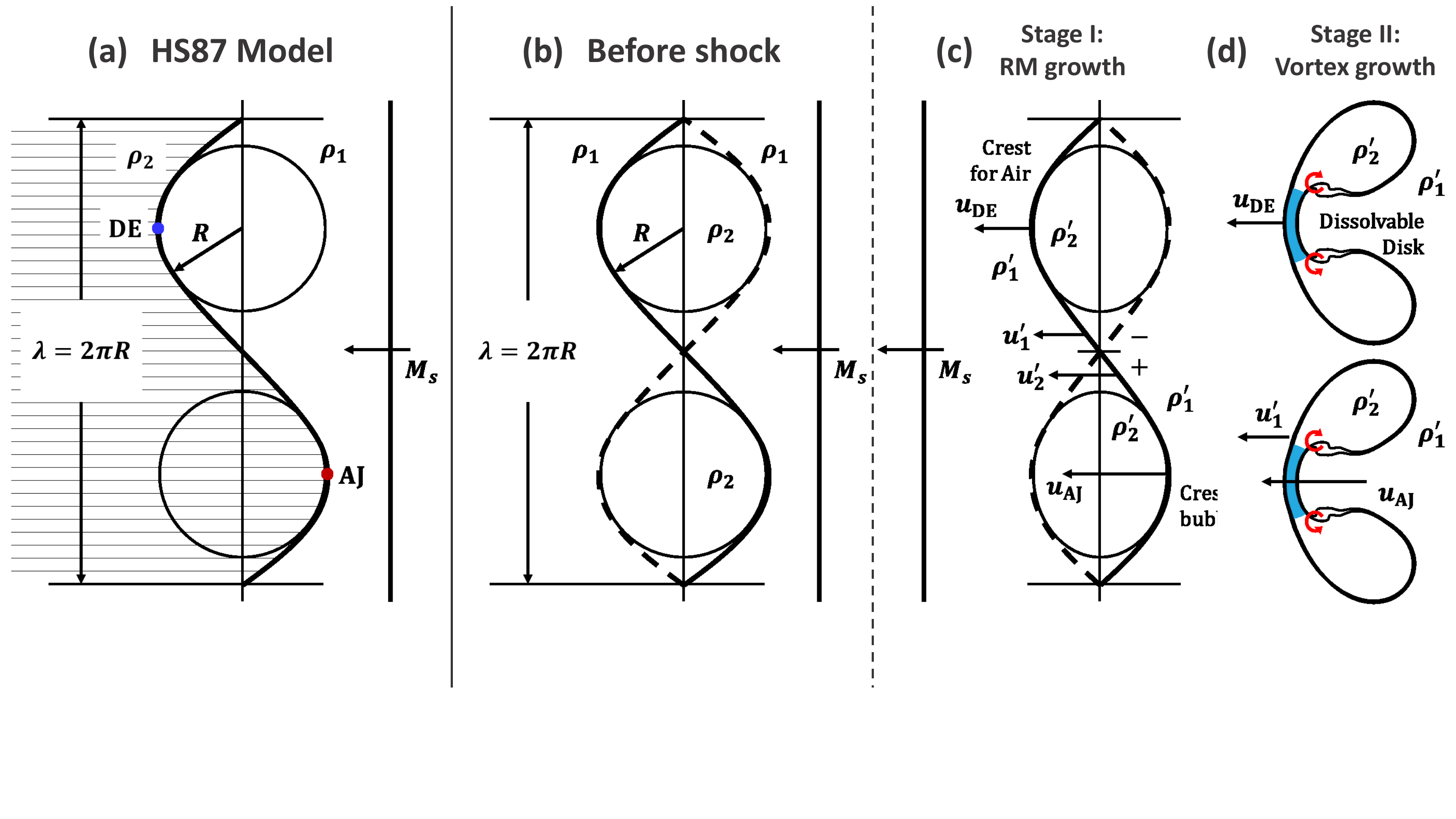}\\
  \caption{The illustration of two-stage growth mode for structure kinetics of SBI. HS model \cite{haas1987interaction} is given in (a) for reference. }\label{fig: illu of modi velo model}
\end{figure}
As shown in Table.\ref{tab: HS model} and \ref{tab: RS model}, we can find that air jet prediction of HS model is fairly well for all Mach number, which means that the incipient formation of shocked bubble still follows the rule of Richtmyer-Meshkov instability at least for air jet. Moreover, for low Mach number, RS model can characterize the vortex propagation under the correct prediction of air jet velocity such as in Ma=1.22 in Table.\ref{tab: RS model}.
However, HS model seems to fail to predict DE velocity in SBI and RS model fails to predict vortex motion in high Mach number.
The fundamental reason may be explained as that shock bubble interaction is a highly non-linear complex system in which initial state and final state are at completely different status.
Either of two schools of structure kinetics can solely explain the whole evolution behavior of shock bubble interaction.
Thus, both HS model and RS model should be applied in dialectical way based on the physical formation of different stage in the problem of shock bubble interaction.

Here, we analyze the two-stage growth mode for different structures in SBI as shown in right part of Fig.\ref{fig: illu of modi velo model}(b-d).
As we can find, SBI is a confined geometry that light gas density is surrounded by ambient air, which is different from single-mode RMI as analogy in HS model. It is more like a gas curtain \cite{vorobieff1998power} than a semi-infinite gas inhomogeneity with different density before shock impact as shown in Fig.\ref{fig: illu of modi velo model}(b).
In fact, immediately after shock, as the first stage, the upstream of bubble can be regarded as the crest of sinuous interface because it locates at the upwind side of the bubble marked as `$+$' in Fig.\ref{fig: illu of modi velo model}(c), which is assumed in HS model. However, for downstream edge, it locates at the rear side `$-$' of shocked bubble, which violates the troughs assumption in HS model.
Because the interface between bubble and shocked air at point of DE is the contact discontinuity, they both share the same velocity in view of one-dimensional shock dynamics \cite{ranjan2011shock}.
Thus, if we stand on the side of the post shocked air, downstream edge of bubble becomes the crest point for shocked ambient air in velocity of $u'_1$ with same curvature as AJ point for shocked bubble in velocity of $u'_2$.
Still following the RM linear growth rate theory, for shocked air, this correlation can be further expressed as:
\begin{equation}\label{eq: DE-2}
  \frac{u_{\textrm{DE}}}{u'_1}=1-\phi Z_cakA^{+}=1+\phi Z_cak\frac{1-\sigma}{\sigma+1}
\end{equation}
From this expression, we also can explains the similar velocity between DE and vortex center from the similar velocity expression referring to Eq.\ref{eq: vor} by invoking $A^{+}=(\sigma-1)/(\sigma+1)$.
Table.\ref{tab: pro model} offers the theoretical values and numerical values of DE structures. Acceptable error within range of $\varepsilon<9\%$ from theoretical value can be found, which validates the proposed model.

As analyzed in RS model, the vortex motion is actually at the second stage growth of structure kinetics and is the result of the penetration of air jet formed at the initial stage as shown in Fig.\ref{fig: illu of modi velo model}(d).
The air jet is deemed as a jet penetration of a dissolvable disk that forms the vortex behind the disk.
This physical model inspires us to combined the well-posed AJ model from linear growth rate theory of RMI to the RS vortical motion model.
The air jet kinetic energy is absorbed by vortex, which motivates the vortex motion \cite{taylor1953formation}:
\begin{equation}\label{eq: vor-2}
  u_{\textrm{V}}-u'_1=\frac{2}{\pi^2}(u_{\textrm{AJ}}-u'_1)
\end{equation}
by a coefficient $2/\pi^2$ for an impulsive motion of an infinite lamina \cite{rudinger1960behaviour}.
Rearranging Eq.\ref{eq: vor-2} by introducing Eq.\ref{eq: AJ}, we obtain:
\begin{equation}\label{eq: vor-3}
  \frac{u_{\textrm{V}}}{u'_1}=1+\frac{2}{\pi^2}\left(\frac{u_{\textrm{AJ}}}{u'_1}-1\right)=
  1+\frac{2}{\pi^2}\left[\frac{u'_2}{u'_1}(1-\phi Z_cakA^+)-1\right]
\end{equation}
From Table.\ref{tab: pro model}, it can find that the proposed model for vortex motion largely ameliorate the theoretical prediction comparing with numerical results especially for higher shock Mach number with error $\varepsilon<3\%$.
\begin{table*}
\begin{ruledtabular}
\begin{center}
\begin{tabular}{lllllll}
\multicolumn{1}{c}{Ma} & DE[Eq.\ref{eq: DE-2}]   & DE{[}meas{]} & $\varepsilon\%$   & Vortex[Eq.\ref{eq: vor-3}]  & Vortex{[}meas{]} & $\varepsilon\%$   \\ \hline
1.22                   & 0.35 & 0.38         & -6.98 & 0.32 & 0.34          & -5.74 \\
1.445                  & 0.52 & 0.56         & -7.51 & 0.49 & 0.53          & -6.79 \\
1.8                    & 0.63 & 0.69         & -8.62 & 0.64 & 0.64          & 0.11  \\
2.4                    & 0.70 & 0.75         & -7.23 & 0.75 & 0.73          & 2.08  \\
3                      & 0.72 & 0.79         & -8.00 & 0.80 & 0.79          & 0.88  \\
4                      & 0.74 & 0.80         & -6.91 & 0.84 & 0.82          & 2.22
\end{tabular}
\caption{Comparison between dimensionless velocity obtained from the proposed theoretical model and numerical result (measured from Fig.\ref{fig: 2d velo}) for downstream edge structure and vortex structure. Error of theoretical value is also given.}
\label{tab: pro model}
\end{center}
\end{ruledtabular}
\end{table*}

In order to further validate the proposed two-stage vortical propagation model, we collected the velocity recorded in literature in SBI. Detailed analysis and comparison can be found in Appendix.\ref{sec: app2}. General agreement is obtained from the theoretical prediction and measured values in literature.

\section{A physical model for lift-off in oblique shock/jet interaction}
\label{sec: liftoff}
From previous two section, we built the spatial-temporal correlation and an improved two-stage growth model of 2D SBI.
By combining this two point, we will finally obtained a physical model of lift-off model for OS/JI based on structure kinetics of SBI.
Referring back to Fig.\ref{fig: 2d3d corr-new}, the slope of any structures in coordinate $Ox'y'z'$ satisfy:
\begin{equation}\label{eq: 2d3d-velocity}
  \frac{dY'/dX'}{dx_{2D}/dt}=\frac{tan\beta}{Ma_{2D}\cdot c_1}=\frac{d\widetilde{Y}'/dX'}{\widetilde{u}}
\end{equation}
Here, $\widetilde{u}$ can be velocity of any structures in 2D SBI such as AJ ($u_{\textrm{AJ}}$), DE ($u_{\textrm{DE}}$) or velocity of vortex motion ($u_{\textrm{V}}$).
Then the slope of $d\widetilde{Y}'/dX'$ of any structures in coordinate of $Oxyz$ is:
\begin{equation}\label{eq: slope trans}
  \frac{dY}{dX}=tan(\widetilde{\beta}-\theta)
\end{equation}
where $\widetilde{\beta}$ is the deviation angle of structure in coordinate of $Ox'y'z'$:
\begin{equation}\label{eq: beta}
  \widetilde{\beta}=arctan\left(\frac{d\widetilde{Y}'}{dX'}\right)=arctan\left(\frac{\widetilde{u}\cdot tan\beta}{Ma_{2D}\cdot c_1}\right)
\end{equation}
This ensures that when structure position is chosen as oblique shock, it leads to $\widetilde{u}=Ma_{2D}\cdot c_1$ and $dY_s/dX=tan(\beta-\theta)$, which satisfies the oblique shock dynamics.

\begin{figure}
  \centering
  \includegraphics[clip=true,trim=10 10 20 10,width=1.0\textwidth]{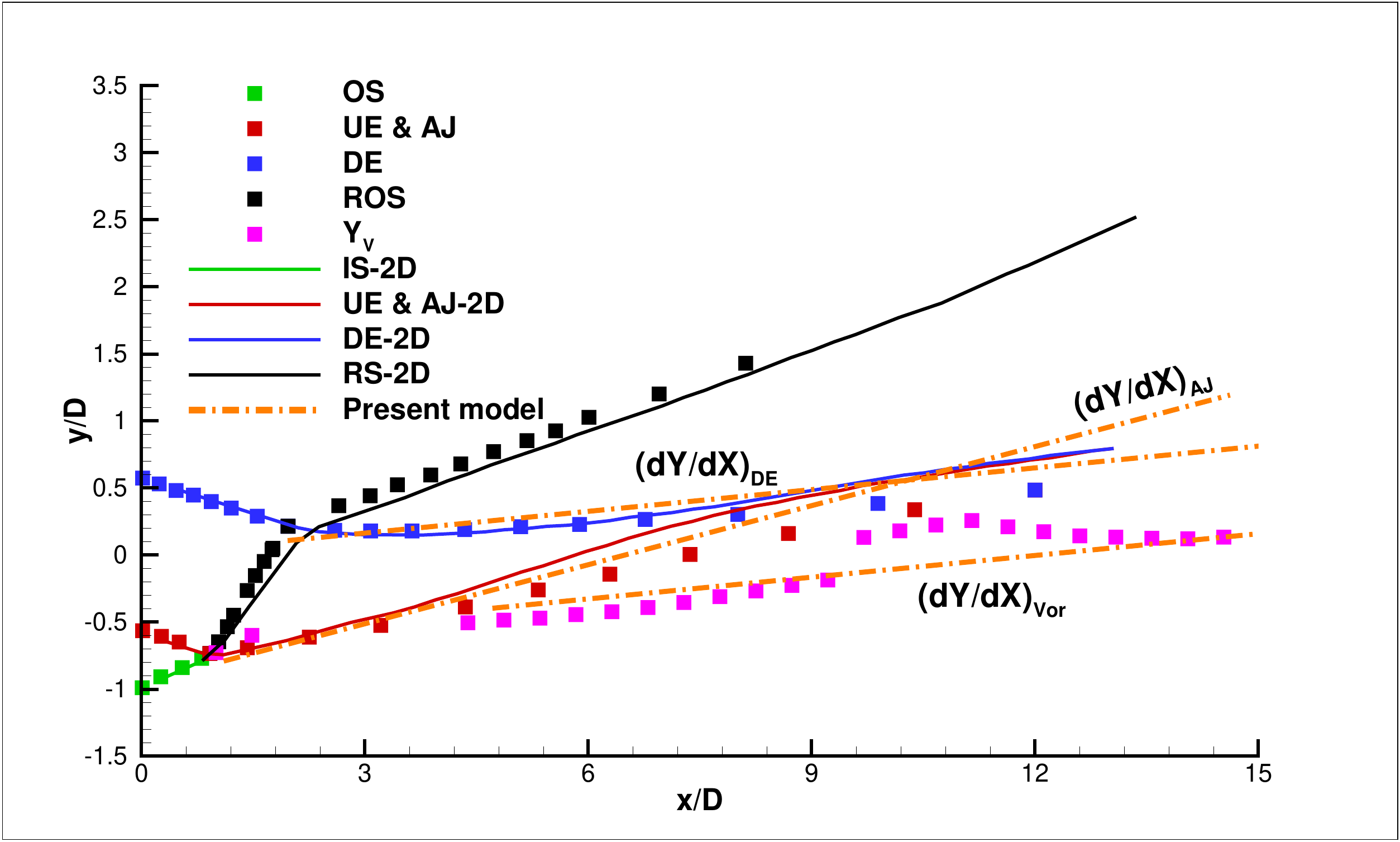}\\
  \caption{Validation of lift-off height model as Eq.\ref{eq: slope trans} by comparing with the results of OJ/SI in present paper.}\label{fig: valid-pres}
\end{figure}
Fig.\ref{fig: valid-pres} shows the prediction of lift-off slope of DE structure, AJ structure and vortex structure by correlating with 2D SBI of Ma$_{2D}$=1.445 case. General agreement is obtained between theoretical value and numerical results in both 2D SBI and 3D OS/JI.

\begin{figure}
   \centering
    \subfigure[Conditions of wall mounted ramp injector in Ref.~\onlinecite{waitz1993investigation}.  ]{
    \label{fig: valid-Waitz-1} 
    \includegraphics[clip=true,trim=30 75 30 0,width=1.0\textwidth]{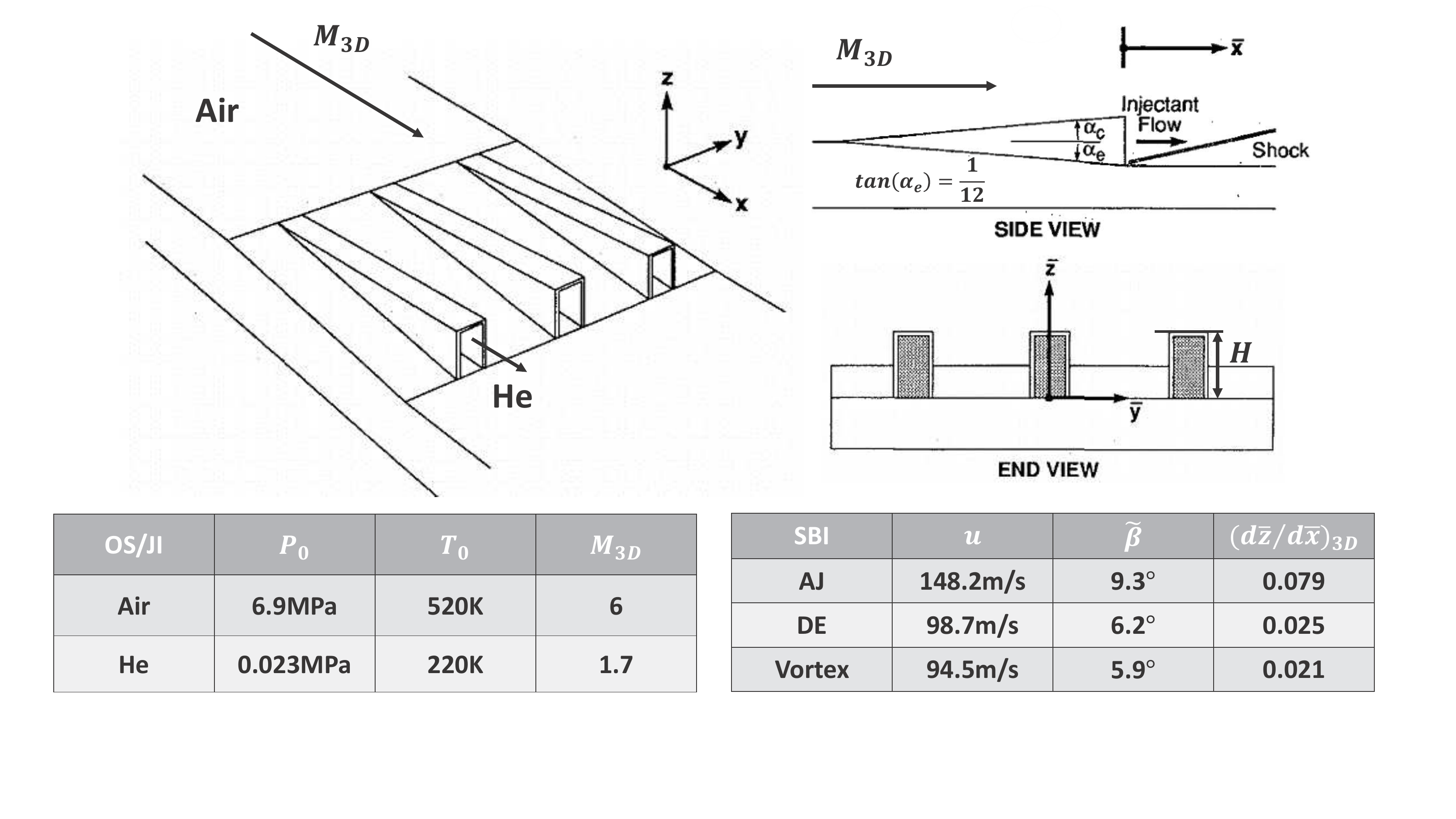}}
    \subfigure[Lift off height of Helium jet.]{
    \label{fig: valid-Waitz-2} 
    \includegraphics[clip=true,trim=10 10 20 10,width=1.0\textwidth]{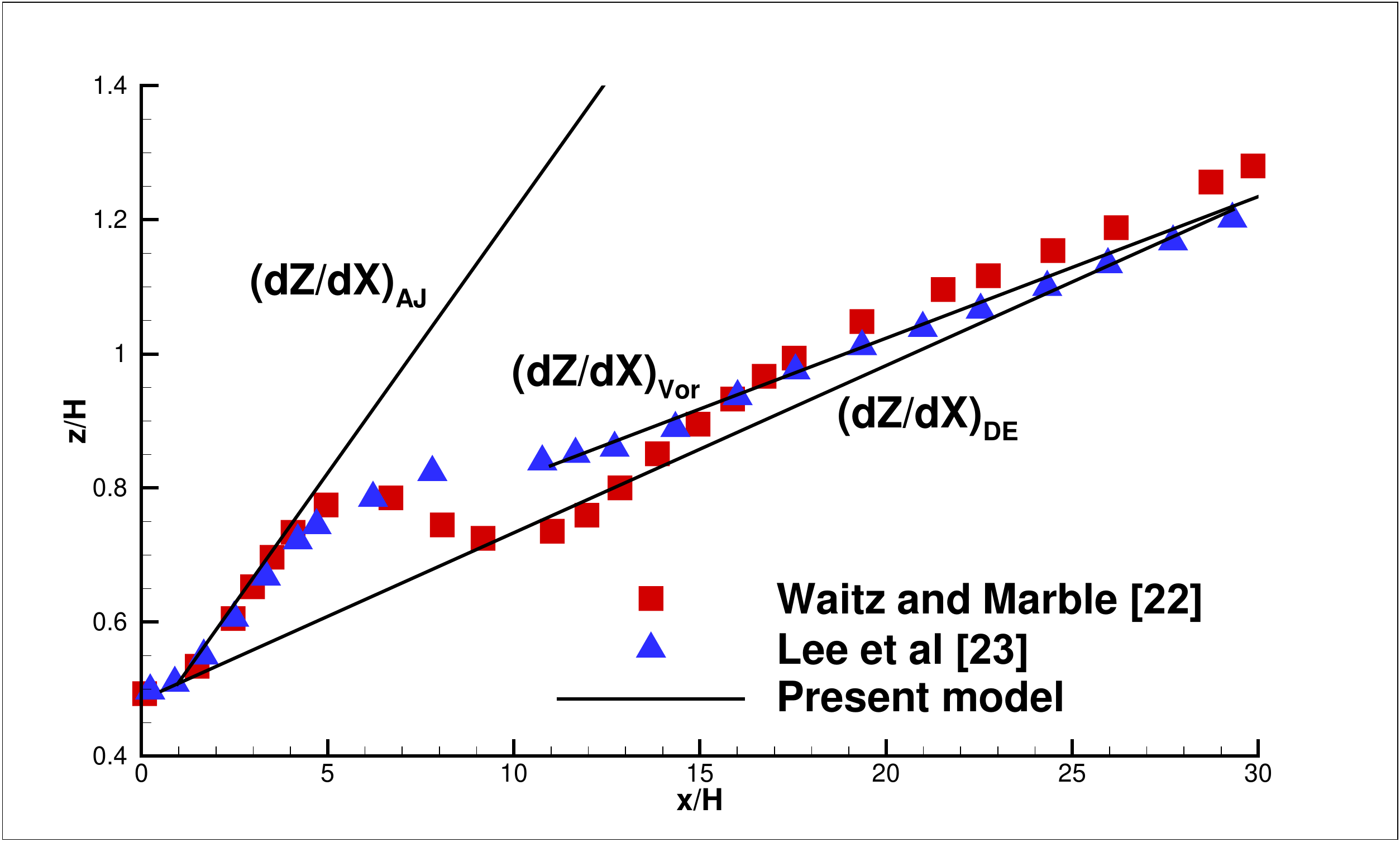}}
    \caption{Validation of lift-off height model as Eq.\ref{eq: slope trans} by comparing with the experimental/numerical results \cite{waitz1993investigation,lee1997computational}.}\label{fig: valid-Waitz}
\end{figure}
Next, this lift-off model shows its effectiveness in predicting the lift-off of fuel jet in a model wall-mounted injector \cite{waitz1993investigation}. This geometry is further studied by numerical method in detail \cite{lee1997computational}, which offers the mutual-validation results.
The thorough experimental conditions are summarized in Fig.\ref{fig: valid-Waitz-1}. With a hypervelocity of Ma$_{3D}$=6, a Helium jet is injector into the combustor. From the total pressure and total temperature combining with the Mach number, we obtain the density of ambient air of $0.239kg/m^3$ and Helium of $0.018kg/m^3$. The deviation angle for compression ramp to produce is $4.76\deg$ which makes a Ma$_{2D}=1.347$. Then using the velocity model proposed, we get the structure position slope angle $\widetilde{\beta}$ summarized in Table inserted in Fig.\ref{fig: valid-Waitz-1}.
Through normalizing the lift-off height by injector height, we plot the theoretical value obtained from Eq.\ref{eq: slope trans} on the data reported separately from Ref.~\onlinecite{waitz1993investigation} and Ref.~\onlinecite{lee1997computational}. At the upstream of jet, the obvious lift-off matches the structure position model of air jet and at the downstream of jet, it shows convergence to structure of vortex and down edge of bubble. It is interesting to note that the centroid of a shocked bubble recorded in \cite{jacobs1992shock} also shows this characteristic of move fast at first, then slowly at late time.

\begin{figure}
   \centering
    \subfigure[LOH dependence on Ma$_{3D}$.]{
    \label{fig: LOH-1} 
    \includegraphics[clip=true,trim=25 20 50 55,width=.48\textwidth]{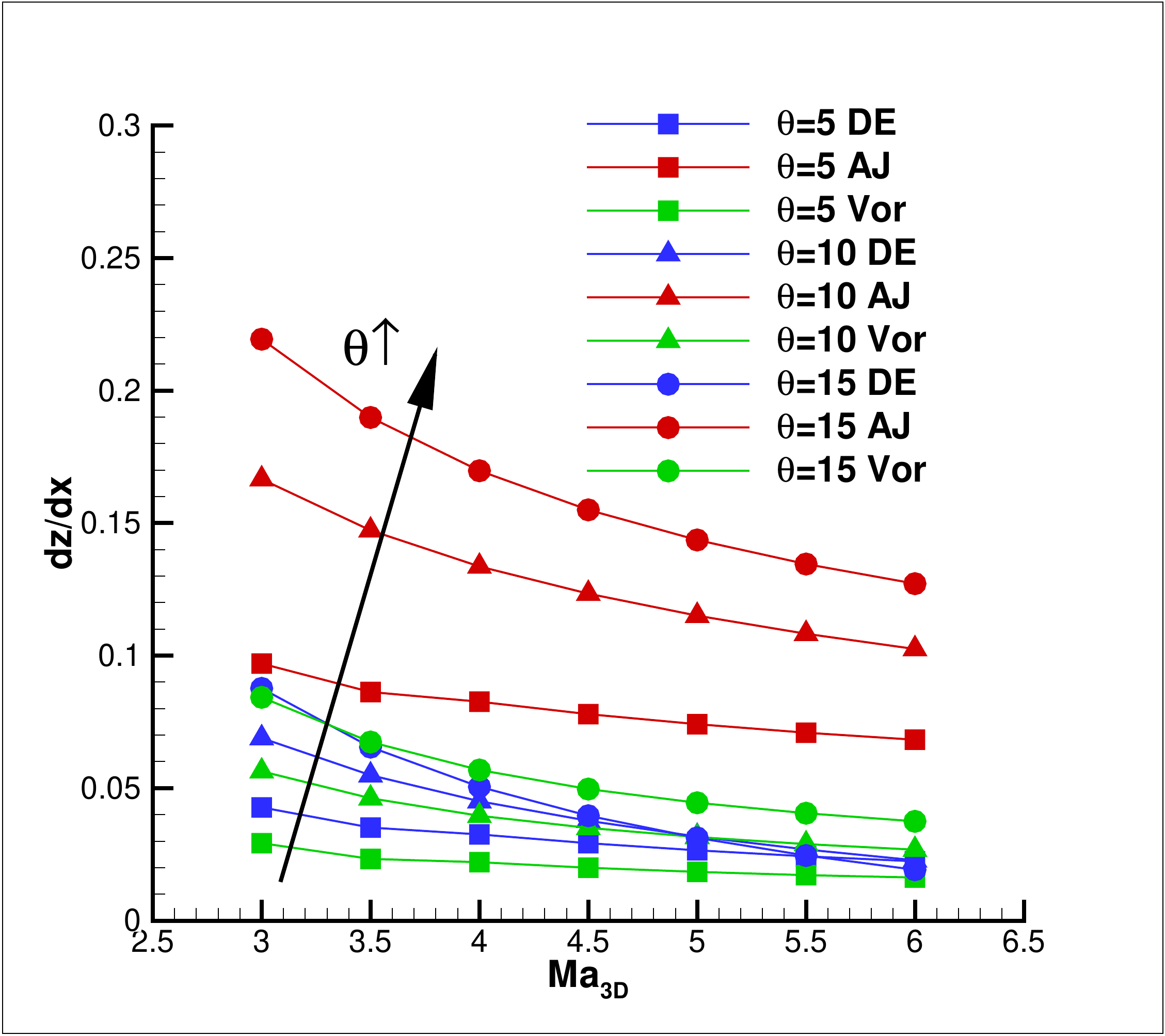}}
    \subfigure[LOH dependence on $\theta$.]{
    \label{fig: LOH-2} 
    \includegraphics[clip=true,trim=25 20 50 55,width=.48\textwidth]{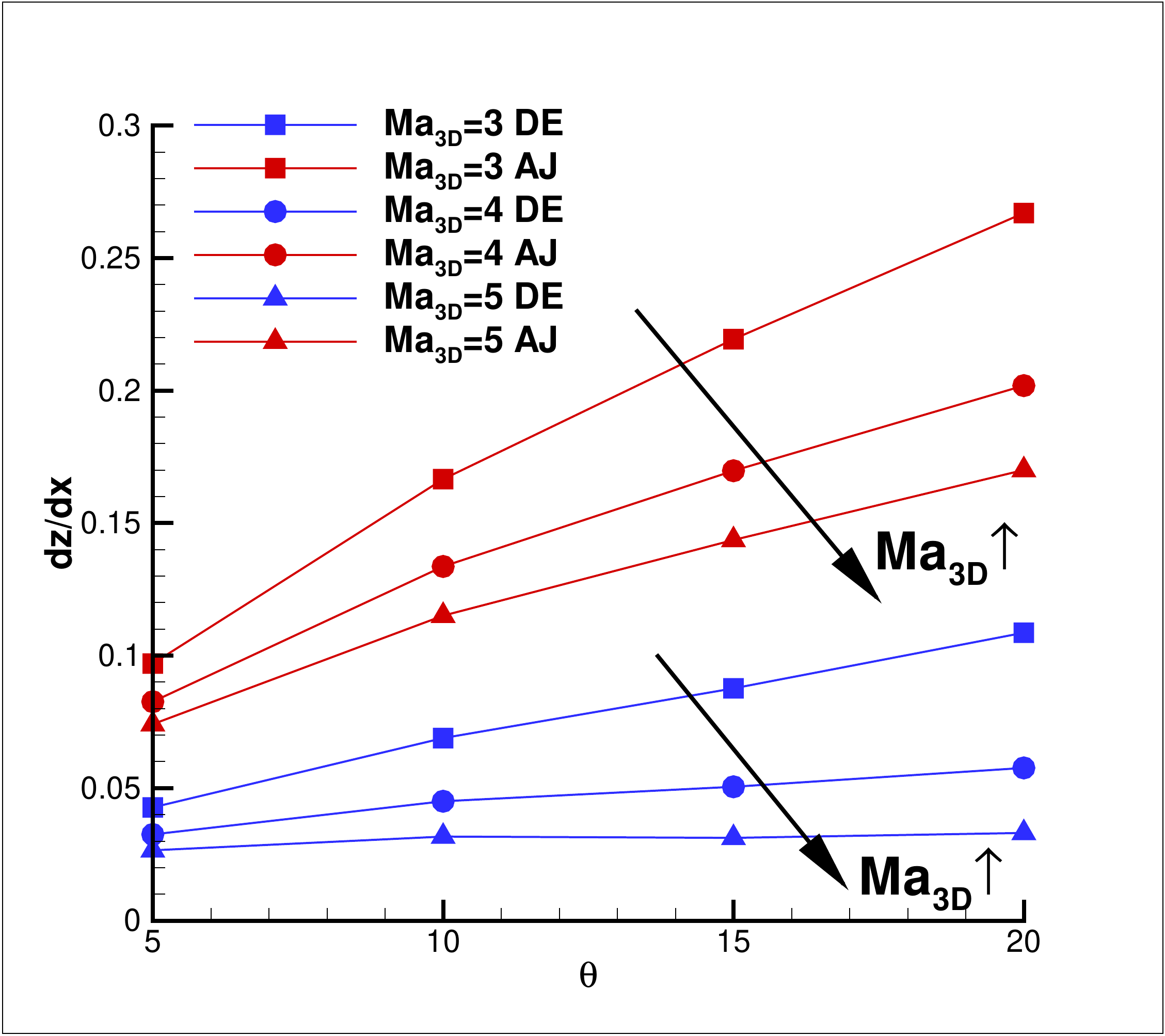}}
    \caption{Liftoff height dependence on inflow Mach number Ma$_{3D}$ and flow compression angle $\theta$. }\label{fig: LOH}
\end{figure}
Finally, we turn our attention to the dependence of lift-off height on inflow Mach number and shock compression angle through theoretical analysis.
For increasing inflow Ma number, the lift-off height decreases steeply especially at high compression angle as shown in Fig.\ref{fig: LOH-1}.
As for lift-off of down edge, a Mach number independence appears, which suggests that for higher inflow Mach number (Ma$>5$), the wall-mounted ramp injector may loses its advantage of increasing lift-off height, while it is relative effective for low inflow Mach number with Ma$<3.5$.

As for deviation angle dependence shown as in Fig.\ref{fig: LOH-2}, increase compression angle shows the nonlinear increase of lift-off height. Facing the total pressure loss formed from oblique shock \cite{menon1989shock}, the synthesized consideration of lift-off height and other performance from a ramp injector is needed.

\section{Concluding remarks}
\label{sec: conclu}
In this paper, the lift-off characteristic from a three dimensional oblique shock/jet interaction is systematically studied.
By standing on the objective frame of oblique shock, a well-posed spatial-temporal correlation is proposed.
For the first time, the detailed analogy between shock bubble interaction and oblique shock/jet interaction is preformed.
The striking similarity between OS/JI and SBI supports the proposition that the lift-off of shock-induced supersonic streamwise vortex is the result of a underlying two-dimensional vortical motion.
Further combining the linear growth rate in Richtmyer-Meshkov instability, suitable for the first stage of structure kinetics of SBI, we raised an improved two-stage vortex motion model that is tested by numerical results in this paper as well as the results from literature to date.
Combining the well-posed spatial-temporal correlation, this kinetics model of SBI leads to a physical lift-off model that shows fairly well prediction of both numerical results of present paper and lift-off height data recorded in previous experimental and numerical studies.

\begin{figure}
  \centering
  \includegraphics[clip=true,trim=0 0 0 0,width=1.0\textwidth]{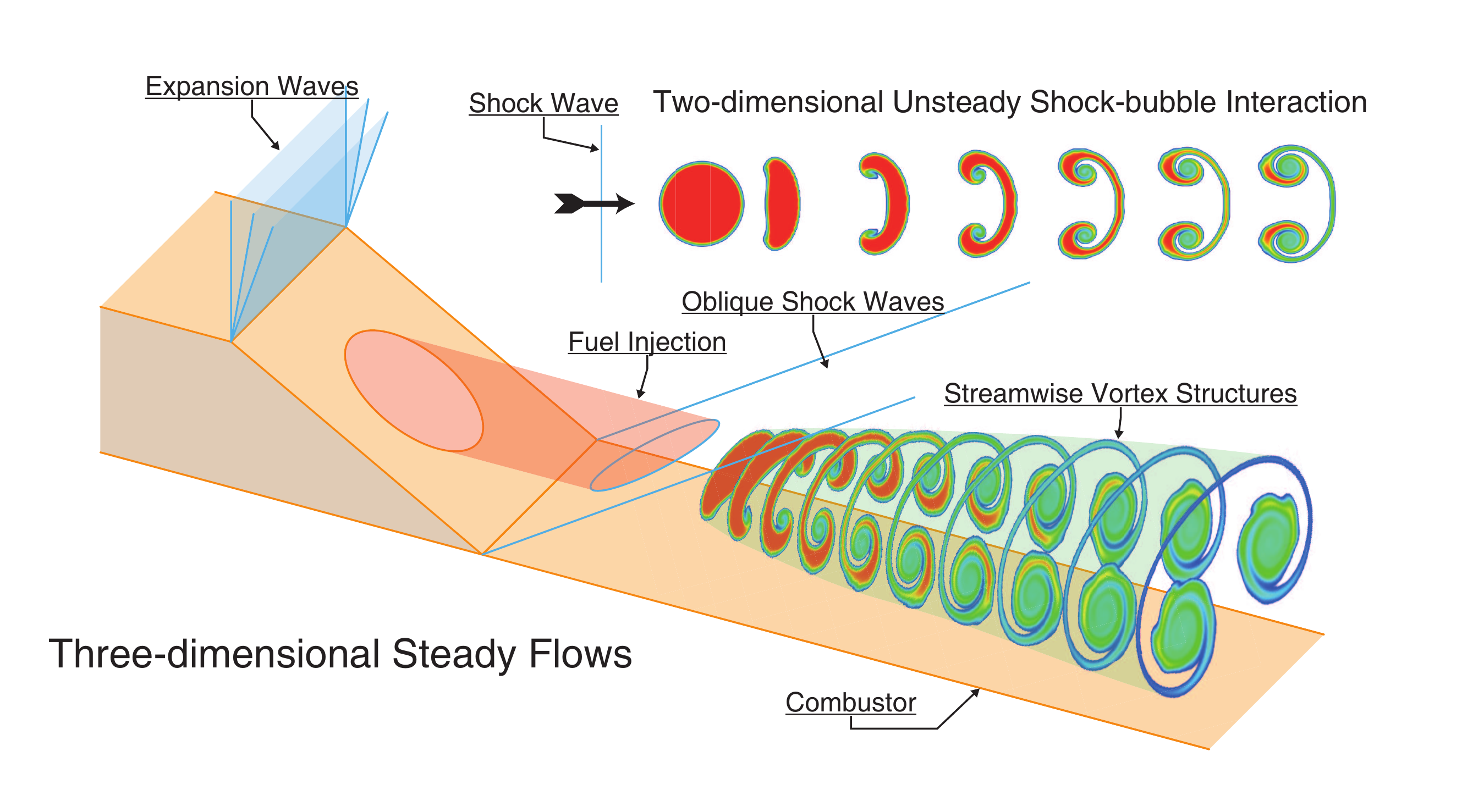}\\
  \caption{The illustration and comparison of the steady three-dimensional oblique shock/jet interaction in scramjet and unsteady two-dimensional shock bubble interaction \cite{Marble1989Progress}.}\label{fig: background}
\end{figure}
In summary, as shown in Fig.\ref{fig: background}, the pioneering idea of Marble \cite{Marble1989Progress} that using oblique shock to enhance mixing and the analogy of SBI and OS/JI further spurs the conversion of theoretical model to application, which also has been employed in relative research \cite{muppidi2006two,maddalena2014vortex}.
This paper provides strong evidence that the extensive research of understanding and controlling method for perturbation growth of RMI can be extended to understand and control the mixing performance of injector in scramjet.
Moreover, the results reported in this paper may pave the new way for future work relating to quantifying the three-dimensional Kelvin-Helmholtz instability on turbulent mixing \cite{dimotakis2000mixing}, combustion \cite{diegelmann2016pressure} and to the preliminary design of scramjet combustor \cite{li2020fuel}.

\begin{acknowledgments}
 This work is financially supported by the the National Natural Science Foundation of China (NSFC) under Grants No. 91441205, No. 91941301 and National Science Foundation for Young Scientists of China (Grant No.51606120). Besides, the support from the Center for High Performance Computing of SJTU for providing the super computer $\pi$ is faithfully acknowledged.
 This work benefited from the three-dimensional data-processing by Yuxuan Li, Chunhui Tang and fruitful discussions with Linying Li.
\end{acknowledgments}

\appendix
\section{Mesh independence study for OS/JI}
\label{sec: app1}
\begin{figure}
  \centering
  \includegraphics[clip=true,trim=0 50 0 25,width=1.0\textwidth]{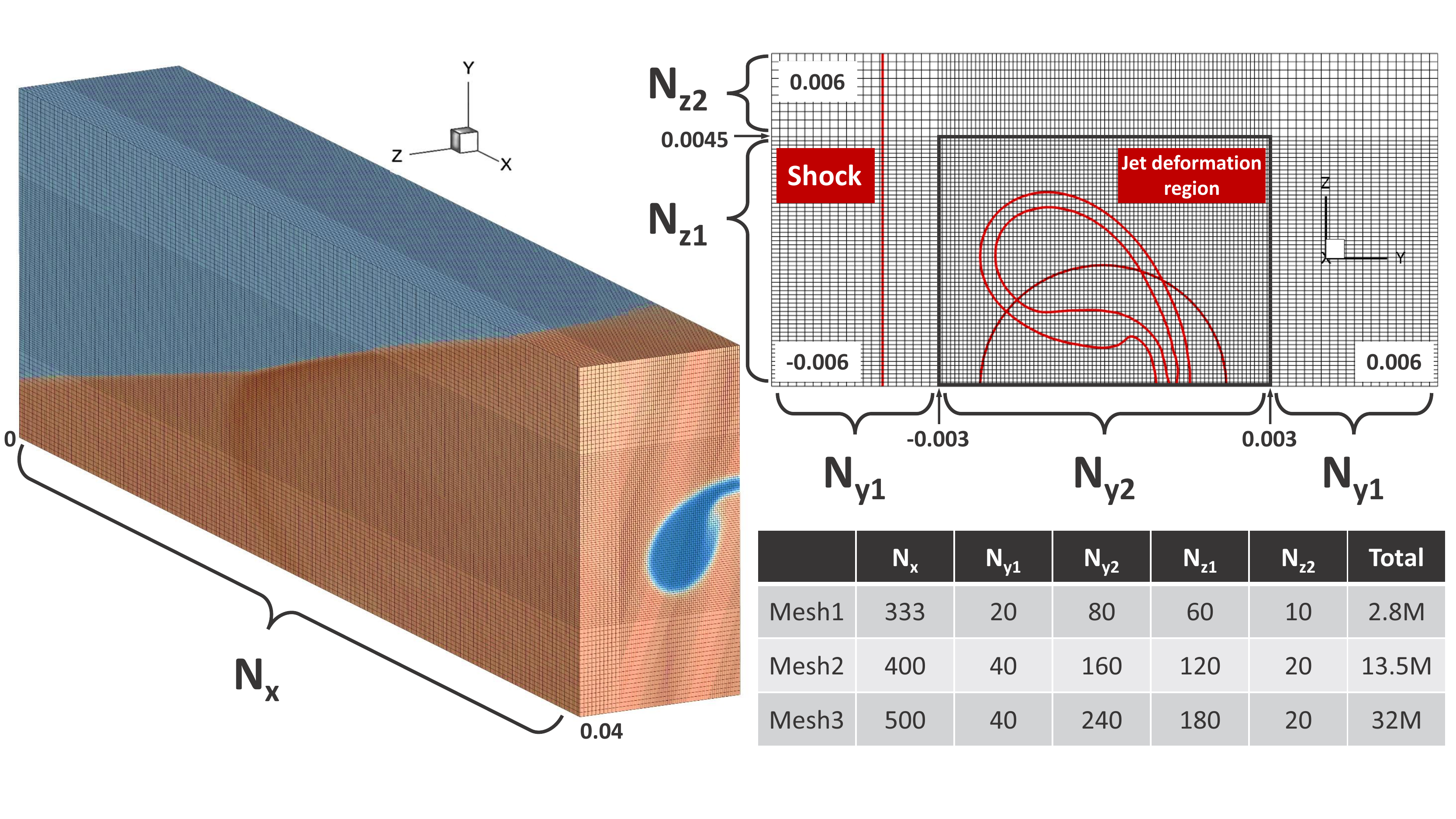}\\
  \caption{Three meshes of mesh independence study of OJ/SI.}\label{fig: App-0}
\end{figure}
Here, the grid dependence study of three dimensional oblique shock jet interaction is investigated. In order to reduce the computational burden in the mesh independence study, direction of streamwise ($x$ direction) is reduced to $0.04m$.
Fig.\ref{fig: App-0} shows the detailed mesh information of three different resolution. In this paper, the core bubble deformation region concerned is particular refined with fine mesh. Three kinds of mesh resolution is up to 2.8 million cells for Mesh-1, 13.44 million cells for Mesh-2 and 32 million cells from Mesh-3. The resolution for three meshes of $\Delta x\times\Delta y\times \Delta z$ are $120\mu m\times 75\mu m \times 75\mu m$ for Mesh-1, $100\mu m\times 37.5\mu m\times37.5\mu m$ for Mesh-2 and $80\mu m\times 25\mu m\times 25\mu m$ for Mesh-3. The initial conditions and boundary conditions are same as the ones introduced in Sec.\ref{sec: numer}.
\begin{figure}
  \centering
  \includegraphics[clip=true,trim=150 150 120 100,width=1.0\textwidth]{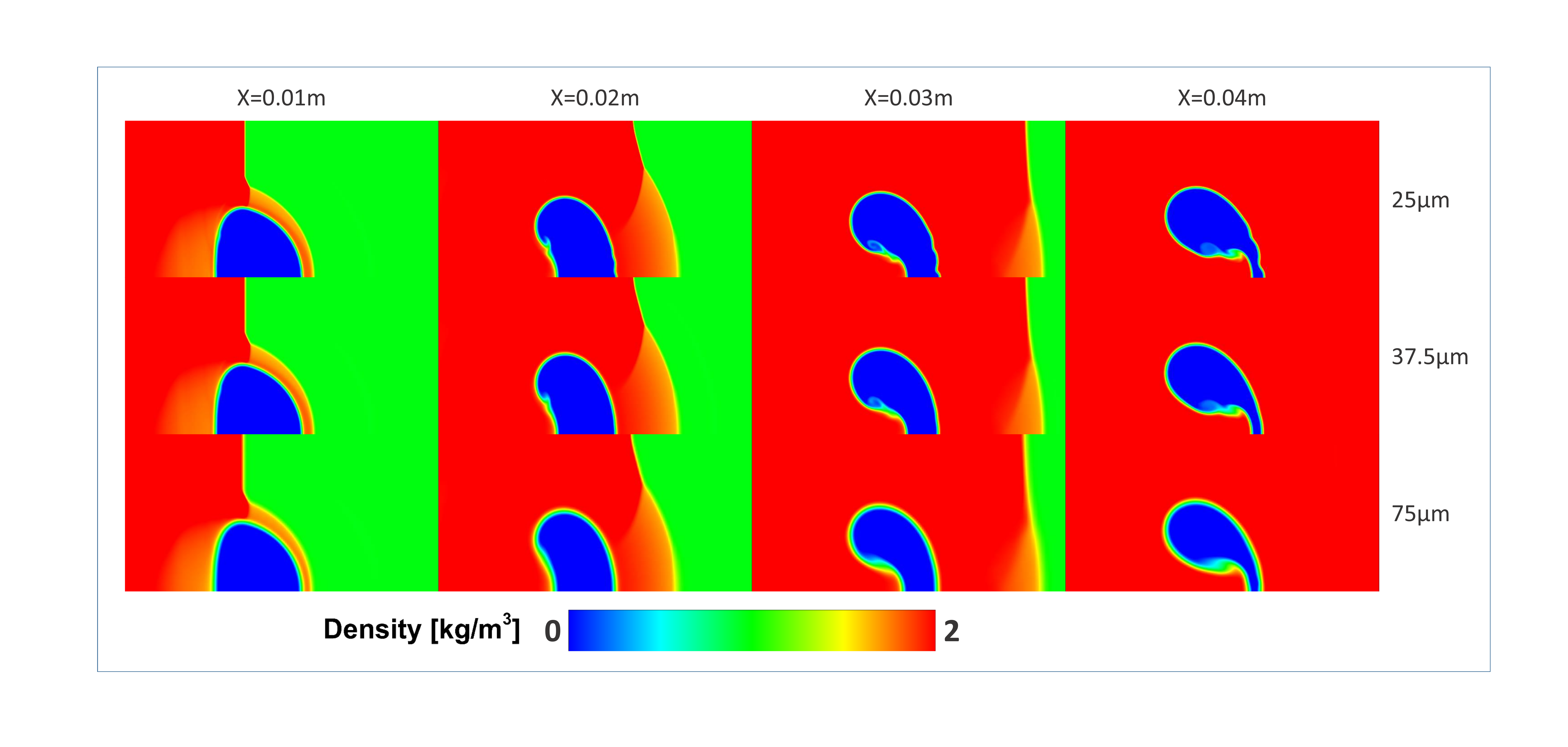}\\
  \caption{Four streamwise section of OJ/SI with different resolution.}\label{fig: App-1}
\end{figure}
\begin{figure}
    \centering
    \subfigure[Total circulation]{
    \label{fig: App-2} 
    \includegraphics[clip=true, trim=10 25 20 10,width=.48\textwidth]{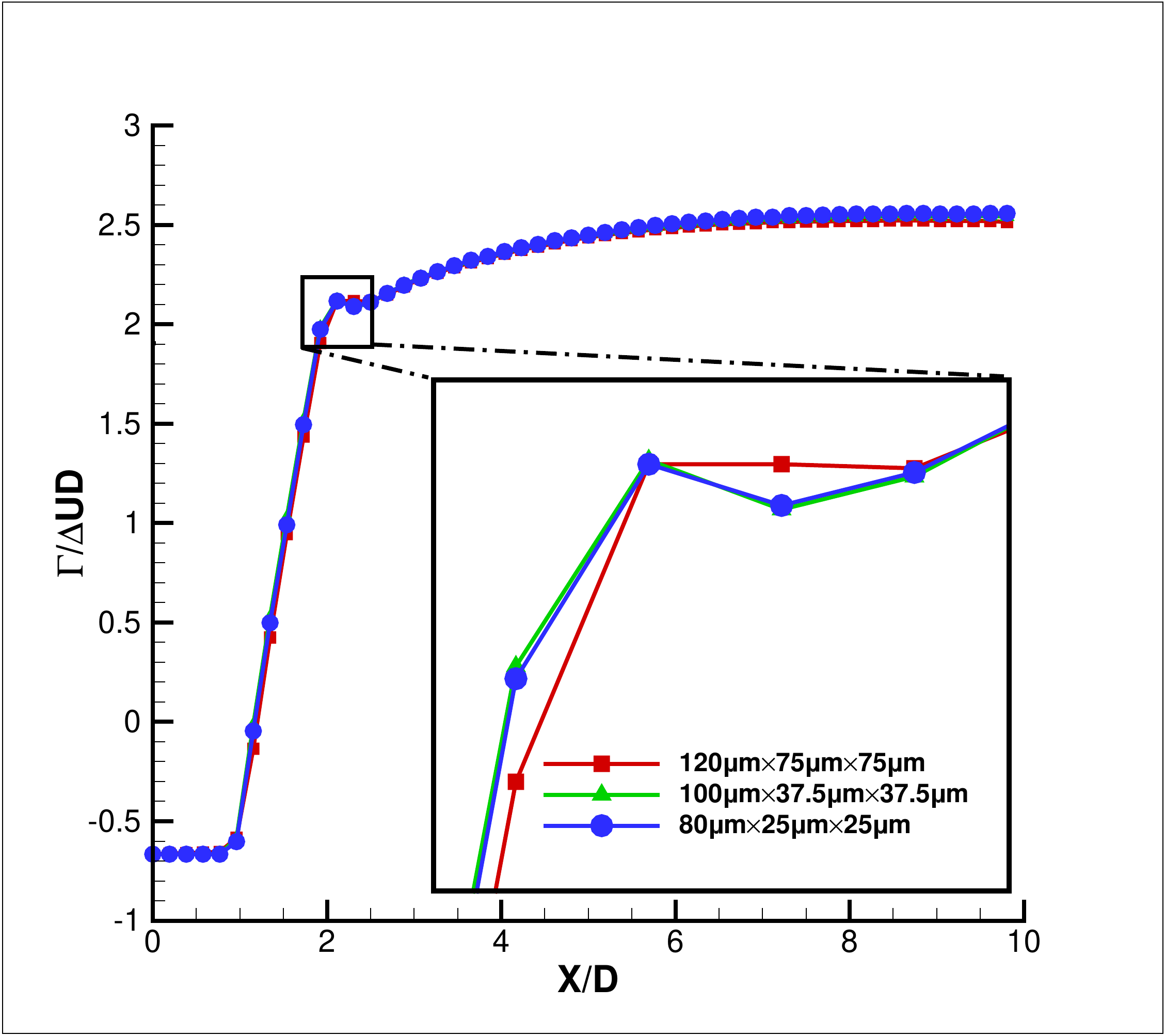}}
    \subfigure[Secondary baroclinic circulation]{
    \label{fig: App-2-1} 
    \includegraphics[clip=true, trim=10 25 20 10,width=.48\textwidth]{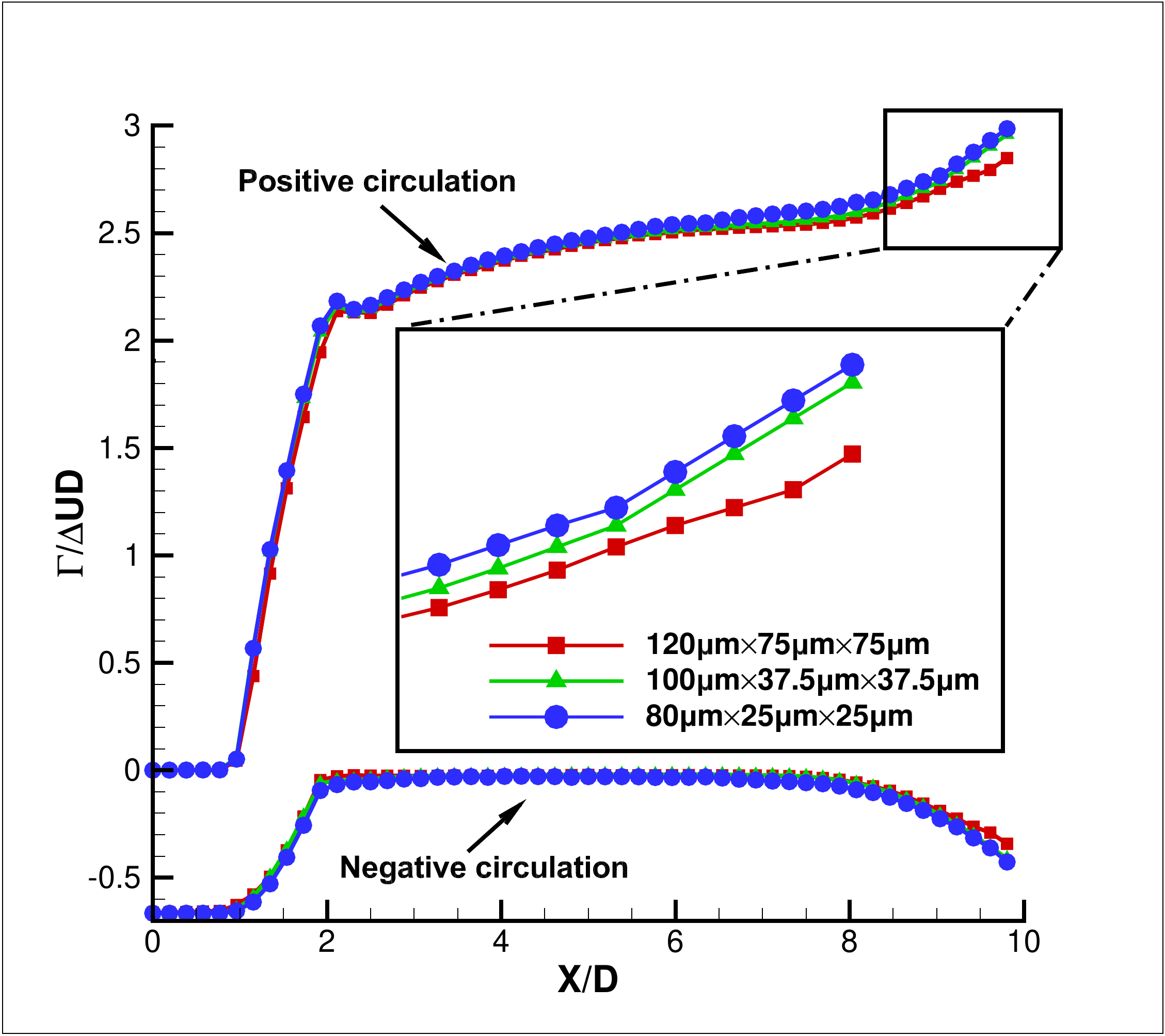}}
    \caption{Comparison of both total circulation and secondary baroclinic circulation of different resolution of OJ/SI (referring to Fig.\ref{fig: circu comp} and Eq.\ref{eq: cirx}). Dimensionless circulation is used where $\Delta U=u'_2-u'_1$ as shown in Table.\ref{tab: initial-condi}.\label{fig: App-2-line} }
\end{figure}
First, the four streamwise cross section flow contours are compared as shown in Fig.\ref{fig: App-1}. The general consistency are obtained for three different resolution. With the increase of the mesh resolution to Mesh-2, the secondary vortex are shown comparing to the low resolution Mesh-1. This secondary vortex is also shown in Mesh-3 of high resolution. Although some instability appears at bridge structure of Mesh-3, good consistency of fine flow structures exists between Mesh-2 and Mesh-3.

Quantitative analysis of cross section circulation of $\omega_x$ also include positive circulation and negative circulation is also compared in Fig.\ref{fig: App-2-line}. The total circulation of three different resolution are compared in Fig.\ref{fig: App-2} and positive and negative circulation produced from secondary baroclinic vorticity of three resolution are compared in Fig.\ref{fig: App-2-1}. Similar to the flow contour, Mesh-2 shows the same trend with Mesh-3 of both total circulation and secondary baroclinic circulation.
In general, the medium mesh of Mesh-2 shows the similarity with the fine mesh of Mesh-3. Considering the reducing the computational burden, the resolution of Mesh-2 are chosen in this study, that is sufficient for capturing lift-off structures correctly in quantitative way.
\begin{table*}
\begin{center}
\begin{tabular}{|l|l|l|l|l|l|l|l|l|}
\hline
\multicolumn{1}{|c|}{Ref} & Ma                                         & At                      & AJ[Eq.\ref{eq: AJ}]                      & AJ{[}meas{]} & DE[Eq.\ref{eq: DE-2}]                      & DE{[}meas{]} & Vor[Eq.\ref{eq: vor-3}]                     & Vor{[}meas{]} \\ \hline
HS87[\onlinecite{haas1987interaction}]\footnote{Experimental study}                      & 1.085                                      & \multirow{4}{*}{-0.69} & 86.49\footnote{$28\%$ air contamination is considered.}                   & 125.00($-30\%$)\footnote{ Measurement error is up to $30\%$ reported in Ref.~\onlinecite{haas1987interaction}.}       & 63.24                   & 69.00($-8.3\%$)        & 60.00                   & 54.85($9.1\%$)         \\ \cline{1-2} \cline{4-9}
HS87[\onlinecite{haas1987interaction}]                      & \multicolumn{1}{c|}{\multirow{3}{*}{1.22}} &                        & \multirow{3}{*}{200.00$^{\text{b}}$} & 230.00($-13\%$)       & \multirow{3}{*}{146.01} & 145.00($0.7\%$)       & \multirow{3}{*}{131.95} & 128.00($3.0\%$)        \\ \cline{1-1} \cline{5-5} \cline{7-7} \cline{9-9}
PB88[\onlinecite{picone1988vorticity}]\footnote{Numerical study}                      & \multicolumn{1}{c|}{}                      &                        &                         & 214.00($-6.6\%$)       &                         & 143.00($2.1\%$)       &                         & 135.00($-2.2\%$)        \\ \cline{1-1} \cline{5-5} \cline{7-7} \cline{9-9}
QK96[\onlinecite{quirk1996dynamics}]$^{\text{d}}$                     & \multicolumn{1}{c|}{}                      &                        &                         & 227.00($-11\%$)       &                         & 146.00($0.01\%$)       &                         & $-$             \\ \hline
Jacobs[\onlinecite{jacobs1992shock}]$^{\text{a}}$                  & 1.093                                      & \multirow{7}{*}{-0.76} & 100.59                  & 88.52($13\%$)\footnote{Named as center upstream edge in Ref.~\onlinecite{jacobs1992shock}.}        & 70.69                   & 67.93($3.9\%$)        & 61.42                   & 57.12($7.5\%$)         \\ \cline{1-2} \cline{4-9}
\multirow{2}{*}{Igra[\onlinecite{igra2018numerical}]$^{\text{d}}$}   & 1.17                                       &                        & 129.69                  & 126.70($2.1\%$)      & $-$                       & $-$            & $-$                       & $-$             \\ \cline{2-2} \cline{4-9}
                          & 1.19                                       &                        & 143.83                  & 140.80($2.0\%$)       & $-$                       & $-$            & $-$                       & $-$             \\ \cline{1-2} \cline{4-9}
\multirow{7}{*}{Yang[\onlinecite{yang1993applications}]$^{\text{d}}$}   & 1.05                                       &                        & 56.52                   & $-$            & 39.73                   & $-$            & 33.95                   & 34.44($-1.4\%$)         \\ \cline{2-2} \cline{4-9}
                          & 1.1                                        &                        & 107.42                  & $-$            & 75.49                   & $-$            & 65.76                   & 66.59($-1.2\%$)         \\ \cline{2-2} \cline{4-9}
                          & 1.5                                        &                        & 399.71                  & $-$            & 280.96                  & $-$            & 273.01                  & 262.61($4.0\%$)        \\ \cline{2-2} \cline{4-9}
                          & 2                                          &                        & 649.99                  & $-$            & 457.98                  & $-$            & 477.34                  & 457.69($4.3\%$)        \\ \cline{2-9}
                          & 1.1                                        & \multirow{2}{*}{-0.47} & 84.07                   & $-$            & 68.57                   & $-$            & 61.02                   & 61.92($-1.4\%$)         \\ \cline{2-2} \cline{4-9}
                          & 2                                          &                        & 573.39                  & $-$            & 468.33                  & $-$            & 461.82                  & 443.85($4.0\%$)        \\ \cline{2-9}
                          & 2                                          & -0.25                  & 511.14                  & $-$            & 458.93                  & $-$            & 449.21                  & 440.14($2.0\%$)        \\ \hline
\end{tabular}
\caption{Validation of structure model by the results of 2D SBI from current literature to date. For convenience, the absolute velocity in unit of $m/s$ is converted for data from different literature. Error of theoretical value $\varepsilon\%$ are given in parentheses.} \label{tab: valid model}
\end{center}
\end{table*}


\section{Validation of two-stage growth mode of structure kinetics in SBI for data from literature}
\label{sec: app2}
In this section, the structure kinetics of shocked bubble introduced in Sec.\ref{sec: 2D SBI} are validated by data for interaction between shock and light cylindrical bubble from the current literature to date. A wide range of Mach numbers and Atwood numbers (referring to pre-shock conditions) are considered as shown in Table.\ref{tab: valid model}. Classical experimental results are obtained from Ref.~\onlinecite{haas1987interaction} and Ref.~\onlinecite{jacobs1992shock}. Several numerical results are also collected \cite{picone1988vorticity,quirk1996dynamics,yang1993applications,igra2018numerical}.
The data from literature tend to be normalized by different ways. For convenience, the absolute velocity in unit of $m/s$ is converted from different literature to compare with values predicted by model.
As for air jet structure, the model from linear RM theory \cite{haas1987interaction,Ding2017On} (see Eq.\ref{eq: AJ}) is slightly lower than numerical and experimental results. For one reason, measurement error is up to $30\%$ as reported in Ref.~\onlinecite{haas1987interaction}. For another reason, although we considered $28\%$ contamination of air uniformly inside bubble, a non-uniform of contamination may occur in real experiment \cite{quirk1996dynamics} which could causes deviation. As for downstream edge of bubble, it can found that general agreement is obtained between theoretical model of Eq.\ref{eq: DE-2} and data from literature. Finally, the improved vortex propagation model of Eq.\ref{eq: vor-3} well captures nearly all cases reported in literature. These results validates the two-stage structure kinetics model of shock bubble interaction introduced in Sec.\ref{sec: 2D SBI}.

\section*{Data Availability Statement}
The data that support the findings of this study are available from the corresponding author upon reasonable request.

\bibliography{mybibfile}

\end{document}